\documentclass[fleqn,usenatbib]{mnras}

\usepackage{newtxtext,newtxmath}

\usepackage[T1]{fontenc}

\DeclareRobustCommand{\VAN}[3]{#2}
\let\VANthebibliography\thebibliography
\def\thebibliography{\DeclareRobustCommand{\VAN}[3]{##3}\VANthebibliography}

\usepackage{graphicx}
\usepackage{amsmath}
\usepackage{longtable}
\usepackage{subfig}

\title[$z=1.89$ Merger]{The MOSDEF Survey: A New View of a Remarkable $z=1.89$ Merger$^{1}$}

\author[J. N. Runco et al.]{Jordan N. Runco,$^{2}$\thanks{E-mail: jrunco@astro.ucla.edu}
Alice E. Shapley,$^{2}$
Mariska Kriek,$^{3,4}$
Michele Cappellari,$^{5}$\newauthor
Michael W. Topping,$^{2,6}$
Ryan L. Sanders,$^{7,8}$
Vasily I. Kokorev,$^{9,10,11}$ 
Sedona H. Price,$^{12}$\newauthor
Naveen A. Reddy,$^{13}$
Alison L. Coil,$^{14}$
Bahram Mobasher,$^{13}$
Brian Siana,$^{13}$\newauthor
Tom Zick,$^{4}$
Georgios E. Magdis,$^{10,11,15}$
Gabriel Brammer,$^{10,11}$
James Aird$^{16}$
\\
$^{1}$Based on data obtained at the W.M. Keck Observatory, which is operated as a scientific partnership among the California Institute of \\ Technology, the University of California,  and the National Aeronautics and Space Administration, and was made possible by the generous  \\ financial support  of the W.M. Keck Foundation.\\
$^{2}$Physics \& Astronomy Department, University of California: Los Angeles, 430 Portola Plaza, Los Angeles, CA 90095, USA\\
$^{3}$Leiden Observatory, Leiden University, PO Box 9513, NL-2300 RA Leiden, The Netherlands \\
$^{4}$Astronomy Department, University of California, Berkeley, CA 94720, USA\\
$^{5}$Sub-Department of Astrophysics, Department of Physics, University of Oxford, Denys Wilkinson Building, Keble Road, Oxford, OX1 3RH, UK \\
$^{6}$Department of Astronomy/Steward Observatory, 933 North Cherry Ave, Rm N204, Tucson, AZ 85721-0065, USA\\
$^{7}$Department of Physics, University of California, Davis, One Shields Ave, Davis, CA 95616, USA\\
$^{8}$Hubble Fellow\\
$^{9}$Kapteyn Astronomical Institute, University of Groningen, P.O. Box 800, 9700AV Groningen, The Netherlands\\
$^{10}$Cosmic Dawn Center (DAWN), Jagtvej 128, DK2200 Copenhagen N, Denmark\\
$^{11}$Niels Bohr Institute, University of Copenhagen, Blegdamsvej 17, DK2100 Copenhagen \O, Denmark\\
$^{12}$Max-Planck-Institut f\"ur Extraterrestrische Physik, Postfach 1312, Garching, 85741, Germany \\
$^{13}$Department of Physics \& Astronomy, University of California, Riverside, 900 University Avenue, Riverside, CA 92521, USA\\
$^{14}$Center for Astrophysics and Space Sciences, University of California, San Diego, 9500 Gilman Dr., La Jolla, CA 92093-0424, USA\\
$^{15}$DTU-Space, Technical University of Denmark, Elektrovej 327, DK-2800 Kgs. Lyngby, Denmark\\
$^{16}$Institute for Astronomy, University of Edinburgh, Royal Observatory, Edinburgh EH9 3HJ, UK
}

\date{Accepted XXX. Received YYY; in original form ZZZ}

\pubyear{2022}

\begin{document}
\label{firstpage}
\pagerange{\pageref{firstpage}--\pageref{lastpage}}
\maketitle

\begin{abstract}
We present a detailed study of a galaxy merger taking place at $z=1.89$ in the GOODS-S field. Here we analyze Keck/MOSFIRE spectroscopic observations from the MOSFIRE Deep Evolution Field (MOSDEF) survey along with multi-wavelength photometry assembled by the 3D-HST survey. The combined dataset is modeled to infer the past star-formation histories (SFHs) of both merging galaxies. They are found to be massive, with log$_{10}(M_{\ast}/M_{\odot}) > 11$, with a close mass ratio satisfying the typical major-merger definition. Additionally, in the context of delayed-$\tau$ models, GOODS-S 43114 and GOODS-S 43683 have similar SFHs and low star-formation rates (log$_{10}$(SFR(SED)/$M_{\odot}/\rm{yr}^{-1}$) $<$ 1.0) compared to their past averages. The best-fit model SEDs show elevated H$\delta_{\rm{A}}$ values for both galaxies, indicating that their stellar spectra are dominated by A-type stars, and that star formation peaked $\sim0.5-1$~Gyr ago and has recently declined. Additionally, based on SED fitting both merging galaxies turned on and shut off star formation within a few hundred Myr of each other, suggesting that their bursts of star formation may be linked. Combining the SFHs and H$\delta_{\rm{A}}$ results with recent galaxy merger simulations, we infer that these galaxies have recently completed their first pericentric passage and are moving apart. Finally, the relatively low second velocity moment of GOODS-S 43114 given its stellar mass, suggests a disk-like structure. However, including the geometry of the galaxy in the modeling does not completely resolve the discrepancy between the dynamical and stellar masses. Future work is needed to resolve this inconsistency in mass.  
\end{abstract}


\section{Introduction} \label{sec:merger_intro}

Within the current $\Lambda$CDM cosmological framework, galaxies grow in mass through both merging events and the smooth accretion of baryons and dark matter. 
Mergers are an important component of galaxy formation models (e.g., \citealt{hop10}), and obtaining empirical constraints on merger rates as a function of galaxy mass and redshift is a current goal for observational galaxy evolution (e.g., \citealt{lot11, cib19, dun19}). 
Merging systems have been observed out to $z\sim6$ \citep{ven17}, approximately $\sim$0.9 Gyr after the big bang. 

Locally (i.e., $z\sim0$), the Sloan Digital Sky Survey (SDSS) has obtained large statistical samples of pre-coalescence galaxy pairs with projected separations of $30-80$ kpc and radial velocity differences of $200-500$ km s$^{-1}$ \citep{ell08,pat11,pat13,scu12,scu15}.
Compared to isolated galaxies at fixed stellar mass, these systems are identified as having gas-phase metallicity depressed by $\sim0.02-0.05$ dex and star-formation rate (SFR) enhanced by $\sim60\%$ out to 30 kpc separation \citep{scu12}. 
Merging systems at $z>1$ are identified through a variety of methods including photometric pairs (e.g., \citealt{wil11, man12, man16, man18}), spectroscopic features (e.g., \citealt{tas14, ven17, dai21}), and visible morphological features such as tidal tails and double nuclei (e.g., \citealt{lof17, kar15}). 

Rest-optical spectroscopy provides a powerful probe of key galaxy properties such as SFR and metallicity, which register the effects of merging interactions. In the last decade, the commissioning of multi-object near-IR spectrographs on large ground-based telescopes has enabled us to obtain large samples of high S/N rest-optical emission-line spectra for galaxies at $z\sim1.5-3.5$. 
Utilizing the MultiObject Spectrometer For Infra-Red Exploration (MOSFIRE; \citealt{mcl12}) instrument on the 10 m Keck I telescope, the MOSFIRE Deep Evolution Field (MOSDEF) survey \citep{kri15} has observed $\sim$1500 galaxies at $1.4\leq z\leq 3.8$ (roughly half of which are at $z\sim2$). 
Previous MOSDEF studies have identified high-redshift mergers using multiple methods. \citet{hor21} identified 55 mergers at $z\sim2$ using the Cosmic Assembly Near-infrared Dark Energy Legacy Survey (CANDELS) morphology catalog \citep{kar15}, and \citet{wil19} spectroscopically identified 30 galaxy pairs at $1.5\lesssim z\lesssim3.5$.

In this study, we present a merger at $z=1.89$ between the galaxies GOODS-S 43114 and GOODS-S 43683, where galaxy IDs are drawn from the 3D-HST survey catalogs of \citet{ske14}. GOODS-S 43114 was targeted by the MOSDEF team, and high-quality $J$-, $H$-, and $K_{\rm{s}}$-band Keck/MOSFIRE spectra were obtained. GOODS-S 43683 was not a MOSDEF target; however, it was serendipitously captured in the MOSFIRE slit given the mask position angle. 
This merger pair has been previously identified by \citet{van10} and is in the CANDELS morphology catalog \citep{kar15}, though it has not been highlighted in previous MOSDEF merger analyses \citep[e.g.,][]{wil19,hor21}. \citet{van10} analyze the {\it HST}
images and low-resolution WFC3 grism spectra of the merger pair, and, in addition, report that both galaxies contain an active galactic nucleus (AGN) based on X-ray luminosity data from \citet{luo08}. 

The goal of this study is to build on the earlier work of \citet{van10} by analyzing the spectroscopic and host-galaxy properties of both galaxies using Keck/MOSFIRE spectra and emission-line corrected SED fitting and comparing with state-of-the-art galaxy merger simulations. 
Specifically, we aim to better understand the galaxy star formation histories (SFHs) and stellar population properties and the evolutionary stage of the merger, by fitting the H$\delta$ absorption feature following the methodology from \citet{zic18} and using it to estimate the second velocity moment and dynamical mass of the system. Our analysis is enhanced relative to previous work on this merger pair based on both the inclusion of higher-resolution rest-optical spectroscopy enabling a dynamical analysis of GOODS-S 43114, and also more systematic SED modeling of both members of the galaxy pair that folds in pan-chromatic observations extending through the mid- and far-IR.
Section \ref{sec:merger_obs_&_methods} provides an overview of the MOSDEF survey and presents the observations, the SED fitting methodology, and derived galaxy properties.
Section \ref{sec:merger_results} presents the results, while Section \ref{sec:merger_discussion} provides a discussion of the key results in the context of current work with galaxy merger simulations. 
Finally, Section \ref{sec:merger_summary} summarizes the key results from this study. 

All emission-line wavelengths are given in the vacuum frame.  Throughout this paper, we adopt a $\Lambda$-CDM cosmology with $H_0$ = 70 km s$^{-1}$ Mpc$^{-1}$, $\Omega_{\rm{m}}$ = 0.3, and $\Omega_\Lambda$ = 0.7.  Also, we assume the solar abundance pattern from \citet{asp09}.

\begin{figure*}
    \includegraphics[width=0.98\linewidth]{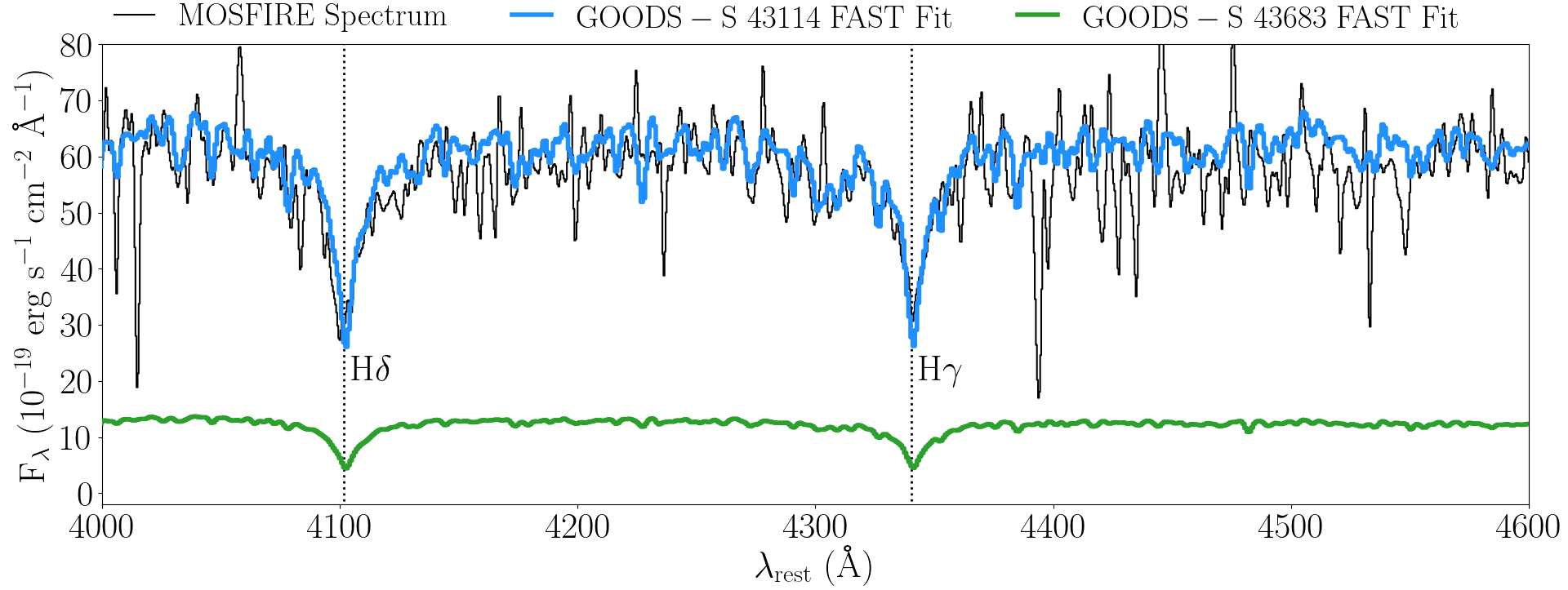}
    \caption{MOSFIRE $J$-band spectrum (black) for GOODS-S 43114 along with the section of the GOODS-S 43114 FAST fit (blue) and GOODS-S 43683 FAST fit (green) that overlaps with the spectra. The H$\delta$ and H$\gamma$ stellar absorption features are labeled. The y-axis is given in units of 10$^{-19}$ F$_{\lambda}$ (i.e., erg s$^{-1}$ cm$^{-1}$ \AA$^{-1}$). The MOSFIRE spectrum has been smoothed to match the resolution of the best-fit FAST model.}
    \label{fig:fast_fit_and_spectra}
\end{figure*}

\begin{figure}
    \includegraphics[width=0.98\linewidth]{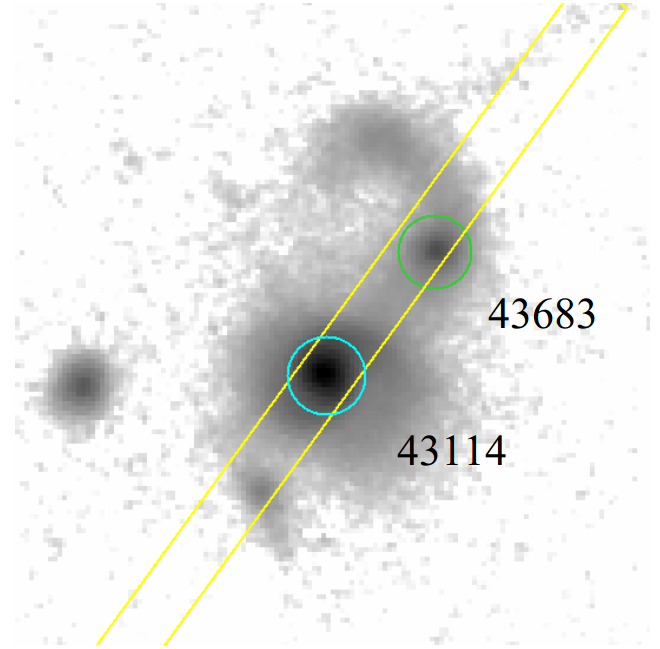}
    \caption{The \textit{HST} WFC3 F160W image from the CANDELS survey with the MOSFIRE slit (yellow), GOODS-S 43114 centroid (blue) and GOODS-S 43683 centroid (green) identified. The image is shown with North up and East to the left.}
    \label{fig:2D_image}
\end{figure}

\begin{figure}
    \includegraphics[width=0.98\linewidth]{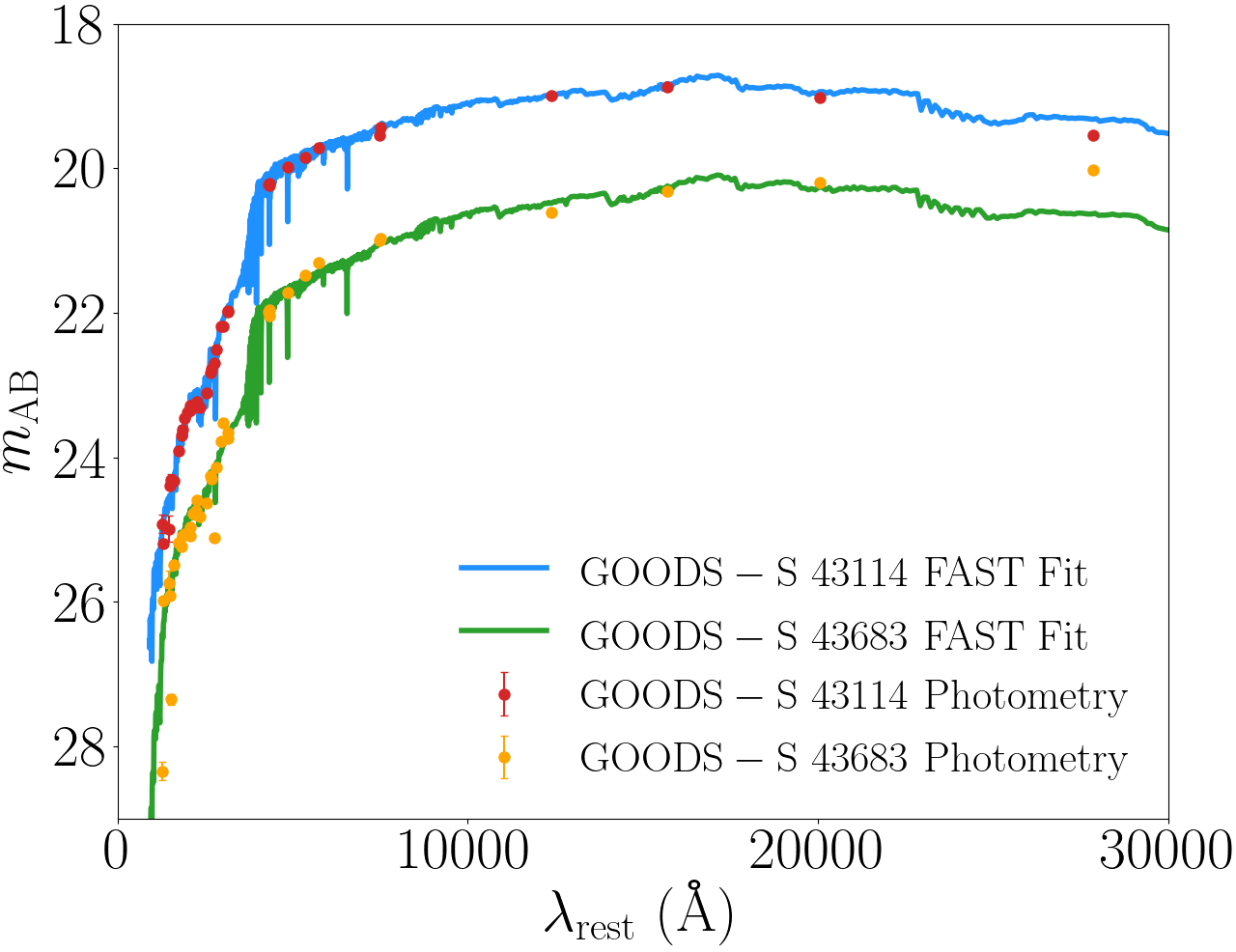}
    \caption{The best-fit FAST models and photometric datapoints for GOODS-S 43114 and GOODS-S 43683, shown in the rest frame. The best-fit FAST model and photometric data are indicated, respectively, with a blue curve and red points for GOODS-S 43114, and a green curve and orange points, respectively, for GOODS-S 43683. The y-axis is given in units of AB magnitude ($m_{\rm{AB}}$).}
    \label{fig:fast_fit_mag_AB}
\end{figure}

\section{Observations \& Methods} \label{sec:merger_obs_&_methods}

\subsection{The MOSDEF Survey} \label{subsec:merger_observations}

Using the MOSFIRE instrument on the 10~m Keck~I telescope, the MOSDEF survey has observed $\sim$1500 galaxies throughout its 48.5-night observing program between 2012-2016. 
MOSDEF galaxies were selected from five well-studied CANDELS and 3D-HST legacy fields \citep{gro11, koe11, mom16} $-$ AEGIS, COSMOS, GOODS-N, GOODS-S, and UDS $-$ and were targeted in three distinct redshift ranges: $1.37 \leq z \leq 1.70$, $2.09 \leq z \leq 2.61$, and $2.95 \leq z \leq 3.80$. These redshift bins were selected to optimise the detection of strong rest-optical emission-lines (e.g., [O~\textsc{II}]$\lambda\lambda$3727,3730, H$\beta$, [O~\textsc{III}]$\lambda\lambda$4960,5008, H$\alpha$, [N~\textsc{II}]$\lambda$6585, and [S~\textsc{II}]$\lambda\lambda$6718,6733) within windows of atmospheric transmission. 
Moderate spectral resolution ($R =$ 3000-3650) was obtained, and the survey is $H$-band (rest-optical) magnitude-limited ($H_{\rm{AB}} =$ 24.0, 24.5, and 25.0, respectively, in the low-, middle-, and high-redshift bins of the MOSDEF sample). 
We use an ABA\textquotesingle B\textquotesingle\space (+1.\textquotedbl 5, 1.\textquotedbl 2, $-$1.\textquotedbl 2, $-$1.\textquotedbl 5) dither pattern to account for detector defects, sky variations, and increase the S/N of the final spectra \citep{kri08}.
For additional MOSDEF observing details, see \citet{kri15}.

\subsection{GOODS-S 43114 \& GOODS-S 43683 Observations} \label{subsec:goodss43114_obs}

GOODS-S 43114 was targeted for MOSDEF spectroscopic observations based on a cataloged spectrosopic redshift of $z=2.6087$ from \citet{bal10}. In fact, this cataloged spectroscopic redshift is inconsistent with both the photometric redshift of  of $z=1.9135$ listed in the 3D-HST catalog \citep[see also][]{sch11}, and the grism spectroscopic redshift presented in \citet{van10}, the latter of which we were unaware of at the time of observation. The 3D-HST photometric redshift in fact places GOODS-S 43114 outside all of the nominal MOSDEF target ranges. However, within the MOSDEF targeting process,  existing spectroscopic redshifts were given priority in the event that a discrepancy arose between spectroscopic and photometric redshift. Using the erroneous spectroscopic redshift for GOODS-S 43114 led to the fortuitous observation of the unique spectra described here.

GOODS-S 43114 was observed across two nights: January 1 and 2 2016. The integration time was 2 hours in each of the $J$, $H$, and $K_{\rm{s}}$ bands with seeing of 0.67, 0.86, and 0.65 arcseconds, respectively. The spectroscopic redshift derived from our MOSFIRE observations is $z=1.8869$, which is consistent with the photometric redshift in the 3D-HST catalog ($1.9135\pm 0.0395$), and slightly lower than the redshift presented in \citet{van10} ($1.902\pm 0.002$). At this redshift, nebular emission lines such as  [O~\textsc{II}]$\lambda\lambda$3727,3730, H$\beta$, [O~\textsc{III}]$\lambda\lambda$4960,5008, H$\alpha$, [N~\textsc{II}]$\lambda$6585, and [S~\textsc{II}]$\lambda\lambda$6718,6733 fall outside of the $J$, $H$, and $K_{\rm{s}}$-bands. 
However, we did observe the H$\gamma$ and H$\delta$ Balmer absorption lines in the $J$~band, from which we derived a spectroscopic redshift for GOODS-S 43114. The $J$-band MOSFIRE spectrum of GOODS-S 43114 is shown in Figure \ref{fig:fast_fit_and_spectra}. No discernible emission- or absorption-line features were captured in the $H$ or $K_{\rm{s}}$-bands. 
For the observation of this merger pair, the MOSFIRE slit width was 0.7'' (5.89 kpc at $z=1.89$) and the FWHM of the profile used for optimal extraction was 1.1'' (9.26 kpc at $z=1.89$).

Figure \ref{fig:2D_image} shows the $HST$ WFC3 F160W 2D image of GOODS-S 43114 and GOODS-S 43683, including identifications of the centroids of the galaxies and the position of the MOSFIRE slit. We measure that the centroids of the two galaxies are 2.135 arcseconds (18 kpc) apart, which agrees with \citet{van10}. 
As previously stated, GOODS-S 43683 was not specifically targeted as part of the MOSDEF survey; however, as Figure \ref{fig:2D_image} shows, it was captured serendipitously in the MOSDEF slit with GOODS-S 43114 given the slit position angle of $-40$ degrees E of N. Unfortunately, due to the the adopted ABA\textquotesingle B\textquotesingle\space dither pattern and our method of using dithered exposures for sky subtraction \citep{kri15}, we cannot detect the spectrum of GOODS-S 43683. Specifically, at a separation of 2.135 arcseconds, the positive continuum of GOODS-S 43683 overlaps with the negative sky-subtraction residual of GOODS-S 43114 offset towards the top of the slit.

GOODS-S 43114 and GOODS-S 43683 have been identified as a merger pair in \citet{van10} and the CANDELS morphology catalog \citep{kar15}. The latter classifies mergers and other galaxy features based on the visual inspection. For GOODS-S 43114 (GOODS-S 43683), 100\% (66\%) of people classified it as having ``any interaction'' while 66\% (100\%) claimed that tidal arms exist. 
GOODS-S 43114 is a class 4 merger, while GOODS-S 43683 is a class 3 merger, both of which are high confidence merger classifications. \citet{van10} also report this galaxy pair as having merger features such as diffuse, tidally-induced spiral arms and tails.

\subsection{SED Fitting \& Derived Properties} \label{subsec:merger_sed_fitting}

We use the SED fitting code FAST\footnote{https://w.astro.berkeley.edu/~mariska/FAST.html} \citep{kri09} to obtain best-fit SEDs and estimates of key galaxy properties for GOODS-S 43114 and GOODS-S 43683. 
With FAST, we adopt the Flexible Stellar Population Synthesis (FSPS) library from \citep{con10} and assume a \citet{cha03} stellar initial mass function (IMF), a \citet{cal00} dust attenuation curve, and delayed-$\tau$ star-formation histories where SFR(SED) $\propto$ $t \times e^{-t/\tau}$ where $\tau$ is the characteristic star-formation timescale and $t$ is the time since the onset of star formation.
We allow $t$, $\tau$, the amount of interstellar extinction ($A_{\rm{V}}$), and stellar mass of the galaxy ($M_{\ast})$ to vary within the models, while fixing the metallicity to 0.019 (defined to be solar metallicity in the \citealt{con10} library). 
For the SED fitting, we fix GOODS-S 43114 and GOODS-S 43683 redshifts to the MOSFIRE spectroscopic redshift obtained for GOODS-S 43114. Given that these galaxies are in a merger with the centroids being only $\sim$18 kpc apart, this redshift assumption for GOODS-S 43683 seems reasonable. \citet{van10} validate our assumption, reporting $z = 1.898 \pm 0.003$ for GOODS-S 43683 (they report $z = 1.902 \pm 0.002$ for GOODS-S 43114). 
In addition, fitting the photometry with the code \texttt{EAZY}\footnote{http://www.astro.yale.edu/eazy/?home} \citep{bra08} yields a photometric redshift of 1.899. 
Along with the best-fit SED and an estimate of $M_{\ast}$, SFR(SED), $A_{\rm{V}}$, and $t/\tau$, we estimate sSFR(SED) from SFR(SED) and $M_{\ast}$. 
UVJ colors are measured from the best-fit SED using the IRAF \citep{tod86, tod93} routine \textit{sbands}.

We use broadband photometry drawn from the 3D-HST v4.1 catalogs \citep{ske14} for both GOODS-S 43114 and GOODS-S 43683. A full description of the method used for multi-wavelength photometric measurements is provided in \citet{ske14}.
For GOODS-S 43114, we fit a combination of the broadband photometry and the $J$-band MOSFIRE spectra. Here the MOSFIRE spectra are normalized to match the flux density corresponding to the photometric points in each band. 
Additionally, for the fitting, we ensure that the $R=3300$ MOSFIRE $J$-band spectrum and FAST models (with a native resolution of 2.5\AA in the rest frame, \citealt{joh21}) are matched in spectral resolution. 
For GOODS-S 43683, we only use the photometry in the fitting process because this target was not spectroscopically detected in the MOSFIRE observations.

The best-fit SEDs and photometry are shown in Figure \ref{fig:fast_fit_mag_AB}. For GOODS-S 43683, there is an offset between the photometry (specifically IRAC channels 3 and 4) and best-fit SED due to the contribution of the AGN. As a sanity check, we refit the photometry with FAST without IRAC channels 2, 3, and 4, and find no significant difference in the estimated galaxy properties. Therefore, the presence of the AGN does not bias the SED fit. 
The portion of the best-fit SEDs that overlaps with the MOSDEF $J$-band spectrum is shown in Figure \ref{fig:fast_fit_and_spectra}.

We estimate the 4000 \AA\ break (D$_{\rm{n}}$4000) and the H$\delta$ absorption-line feature (H$\delta_{\rm{A}}$) using the wavelength ranges from, respectively, \citet{wor97} and \citet{bal99}. D$_{\rm{n}}$4000 traces the opacity of stellar atmospheres and increases with metallicity and age. 
H$\delta_{\rm{A}}$ peaks when A-type stars dominate the spectrum, which occurs when a short burst of star formation is followed by rapid quenching. 
Therefore, the combination of these features can reveal information about the star formation timescale and evolutionary phase of a galaxy, as H$\delta_{\rm{A}}$ is sensitive to recently-quenched star formation while D$_{\rm{n}}$4000 is sensitive to age.
We utilize the best-fit FAST models to measure both quantities for both GOODS-S 43114 and GOODS-S 43683. For GOODS-S 43114, it is also possible to estimate H$\delta_{\rm{A}}$ from the spectra; however, part of the D$_{\rm{n}}$4000 feature is blue-ward of the $J$-band spectrum, so we do not use the spectrum to estimate that quantity. The measurement of H$\delta_{\rm{A}}$ from the model (i.e., 7.81\AA) is consistent at the $1-2\sigma$ level with that estimated from  direct integration of the $J$-band spectrum itself (H$\delta_{\rm{A, spectrum}}$ = 8.42$_{0.56}^{0.56}$ \AA). 

We adopt morphological properties and uncertainties from the F160W catalog of \citet{van14}, in particular the half-light radius ($R_{\rm{e}}$), the S\'ersic index ($n$), the axis ratio ($q$), and the luminosity estimated from the S\'ersic fit ($L_{\rm{Ser}}$). \citet{van14} used a single-component S\'ersic profile fit to the two-dimensional light distribution of the F160W band to estimate these properties. The associated methodology for estimating the associated uncertainties on morphological properties is described in detail in \citet{van12}.

We used the Penalized Pixel-Fitting routine (\texttt{ppxf}\footnote{https://pypi.org/project/ppxf/}; \citealt{cap04, cap17}) to estimate the second moment of the line-of-sight velocity distribution within the half-light radius ($\sigma_{\rm{e}}$) for GOODS-S 43114. 
We fit the MOSFIRE $J$-band spectrum using the \citet{vaz10} stellar population library. 
Specifically, we used a subset of 150 single stellar population template spectra with 25 stellar population ages logarithmically spaced between 0.063 and 15.8 Gyr and six metallicities, [M/H]=($-$1.71, $-$1.31, $-$0.71, $-$0.40, 0.00, and 0.22). In the fit, we used additive polynomials of degree four. 
Since the MOSFIRE spectrum is higher resolution than the stellar templates, we corrected the fitted second velocity moment upwards by the quadratic differences of the two resolutions using the following equation:
\begin{equation}
    \sigma_{\rm{e}}^{2} = \sigma_{\rm{obs}}^{2} + \sigma_{\rm{instr\_temp}}^{2} - \sigma_{\rm{instr\_gal}}^{2}
\end{equation}
where $\sigma_{\rm{instr\_temp}}^{2}$ and $\sigma_{\rm{instr\_gal}}^{2}$ are the instrumental dispersions of the templates and MOSFIRE spectrum, respectively, $\sigma_{\rm{obs}}^{2}$ is the observed second moment, and $\sigma_{\rm{e}}^{2}$ is the second velocity moment corrected for the difference in template and spectrum resolutions. 
Note that both the H$\gamma$ and H$\delta$ lines are included in the fit. \citet{tan19} show that the choice of stellar population library and whether or not to mask the Balmer absorption lines does not bias the estimated second velocity moment. 
We computed bootstrap uncertainties on $\sigma_{\rm{e}}^{2}$ by perturbing the MOSFIRE $J$-band spectrum within its error spectrum and randomly changing the polynomial degree for the fit between 1 and 5.

The second velocity moment, $\sigma_{\rm{e}}^{2}$,  is used to estimate the dynamical mass ($M_{\rm{dyn}}$) using the following methodology. First, we estimated the virial $M/L$ ratio using the following equation:
\begin{equation}
\label{eqn:virial_ml_ratio}
    (M/L)_{\rm{vir}} = \frac{\beta(n) \times R_{\rm{e}} \times \sigma_{\rm{e}}^{2}}{L_{\rm{Ser}} \times G}
\end{equation}
where $G$ is the gravitational constant and, as described above, the remaining variables and uncertainties are adopted from the \citet{van14} catalog (i.e., $R_{\rm{e}}$, $n$, and $L_{\rm{Ser}}$). We adopted equation 20 from \citet{cap06} to estimate the virial coefficient ($\beta$($n$)) using the S\'ersic index. We used this $M/L_{\rm{vir}}$ ratio to estimate $M_{\rm{dyn}}$ according to the following equation:
\begin{equation}
\label{eqn:dynamical_mass}
    M_{\rm{dyn}} = (M/L)_{\rm{vir}} \times L_{\rm{FAST}}
\end{equation}
where $L_{\rm{FAST}}$ is the F160W luminosity from the best-fit FAST model.

\subsection{SDSS Comparison Sample} \label{subsec:merger_sdss_comparison_sample}

In Section \ref{sec:merger_results}, we compare the D$_{\rm{n}}$4000 and H$\delta_{\rm{A}}$ values of GOODS-S 43114 and GOODS-S 43683 with those of local galaxies using archival data from SDSS Data Release 7 (DR7; \citealt{aba09}), specifically from the MPA-JHU DR7 release of spectrum measurements\footnote{https://wwwmpa.mpa-garching.mpg.de/SDSS/DR7/}. 
We use the H$\delta_{\rm{A}}$ and D$_{\rm{n}}$4000 values from the catalog where the measured continuum indices in the spectra have been corrected for sky-line contamination using the best-fit model spectrum. 
We restrict the SDSS sample to a redshift range of $0.04 \leq z \leq 0.10$. 
These criteria give us a final local sample of 305,005 galaxies.

\section{Results} \label{sec:merger_results}

\begin{figure*}
    \includegraphics[width=0.49\linewidth]{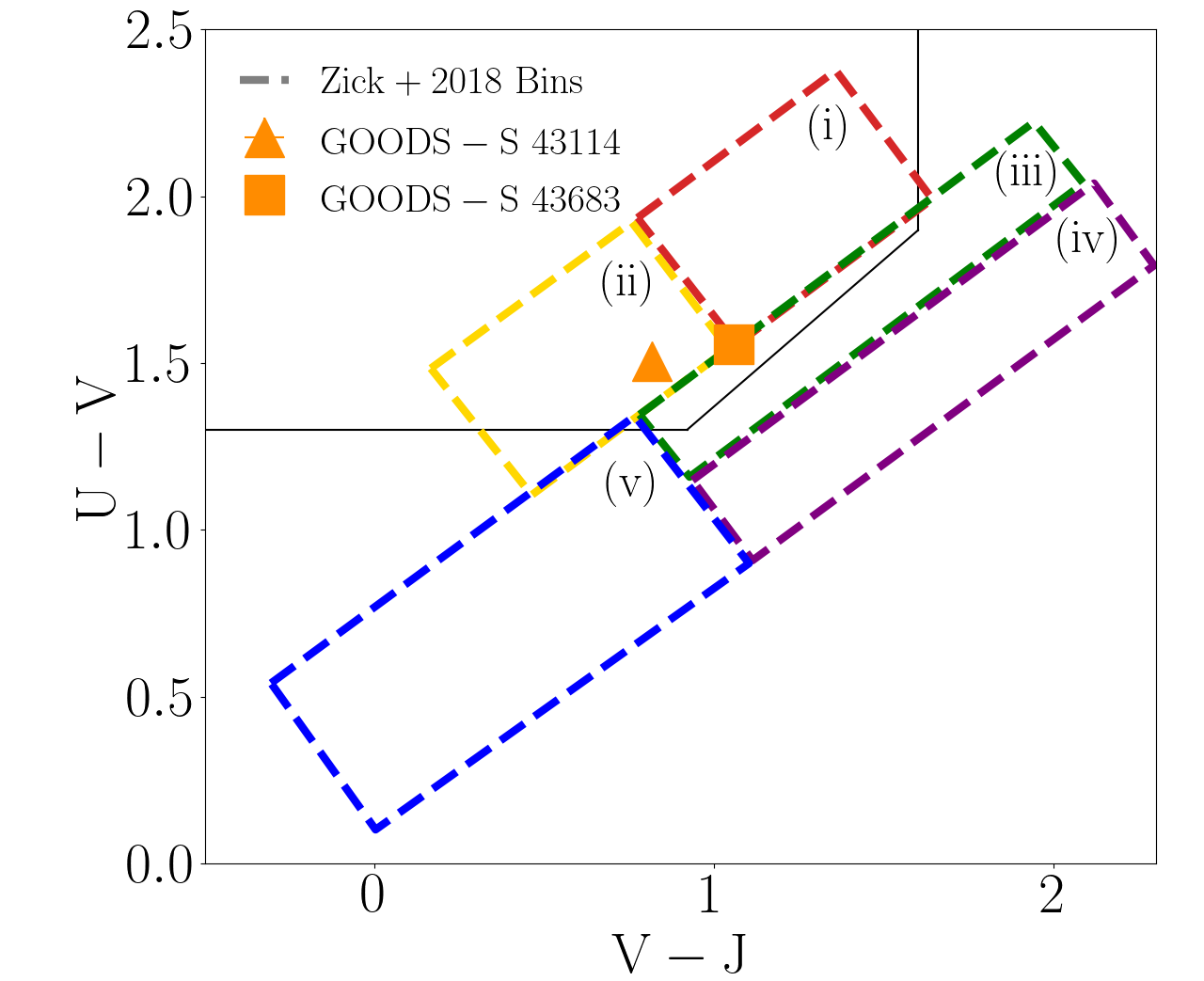}
    \includegraphics[width=0.49\linewidth]{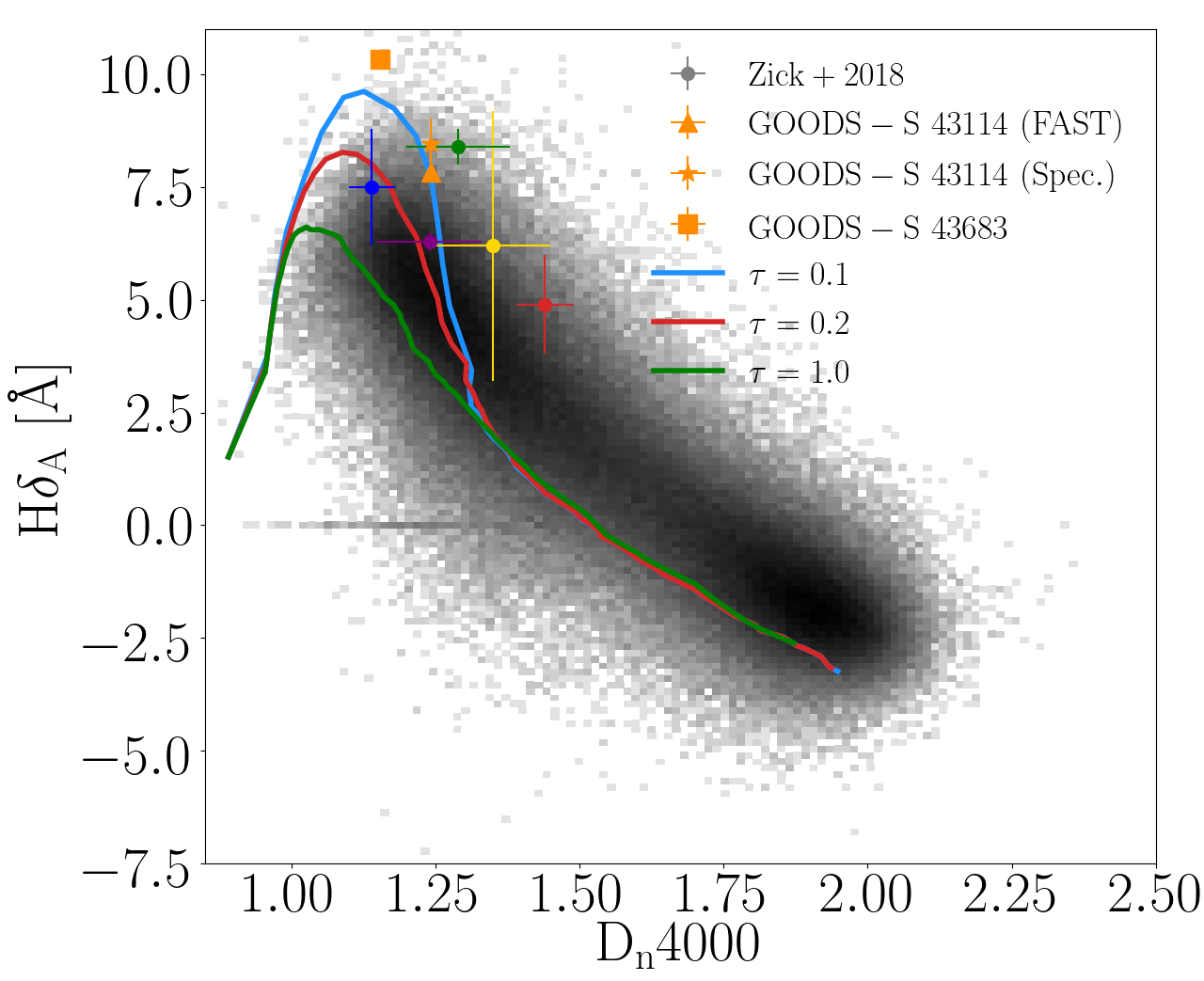}
    \caption{Left: UVJ diagram. GOODS-S 43114 and GOODS-S 43683 are shown by the orange triangle and square, respectively. 
    The dashed boxes identify the five bins of galaxies from \citet{zic18}: quiescent (red), post-starburst (yellow), dusty galaxies with lower sSFRs (green), dust star-forming (purple), and non-dusty star-forming (blue). GOODS-S 43114 falls into the post-starburst bin (i.e., bin (ii)), while GOODS-S 43683 falls at the intersection of bins (i), (ii), and (iii). The black line identifies the quiescent box defined in \citep[][]{wil09}. Right: H$\delta_{\rm{A}}$ vs. D$_{\rm{n}}$4000. The red, yellow, green, purple, and blue circles are color matched to identify the five bins of galaxies shown in the UVJ diagram. The blue, red, and green lines identify three delayed-$\tau$ SFHs, where $\tau$ = 0.1, 0.2, and 1.0 Gyr, respectively. The grayscale 2D histogram indicates local SDSS galaxies. The orange star identifies the H$\delta_{\rm{A}}$ value measured from the MOSFIRE $J$-band spectrum of GOODS-S 43114, while the orange triangle and square have the same meaning as in the left-hand panel. The D$_{\rm{n}}$4000 values for both galaxies are taken from the best-fit FAST models.}
    \label{fig:zick2018_plots}
\end{figure*}

\begin{figure}
    \includegraphics[width=0.98\linewidth]{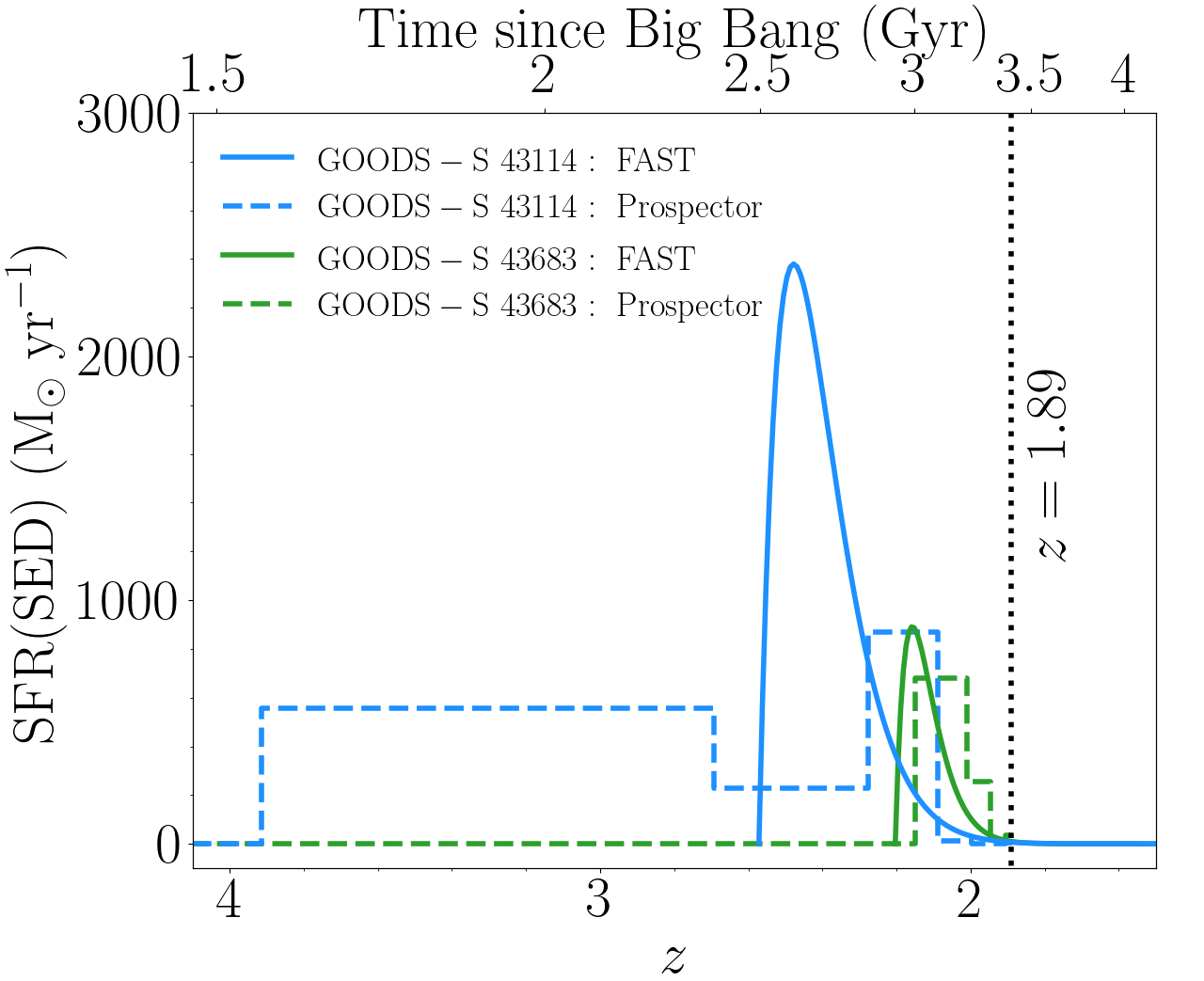}
    \caption{SFR(SED) vs. $z$ displaying the best-fit SFHs assuming a delayed-$\tau$ model for GOODS-S 43114 (blue) and GOODS-S 43683 (green). The black dashed line identifies the epoch of observation (i.e., $z=1.89$). The rapid burst of star formation followed by rapid quenching over the course of $\sim$500 Myr or less gives an approximation of when the first pericentric passage took place.}
    \label{fig:sf_histories}
\end{figure}

\subsection{Galaxy Stellar Populations}

The galaxy properties discussed in Section \ref{subsec:merger_sed_fitting} are shown in Table \ref{tab:fast_and_spec_property_values}. 
The reported 1$\sigma$ uncertainties for best-fit galaxy properties (i.e., $M_{\ast}$, SFR(SED), sSFR(SED), $\tau$, $t/\tau$, and $A_{\rm{V}}$) are estimated from perturbing the photometry and each wavelength element of the MOSFIRE $J$-band spectrum within their uncertainties and refitting 500 times. The same procedure was adopted to estimate uncertainties on properties obtained from the best-fit SED (i.e., UVJ colors, D$_{\rm{n}}$4000, and H$\delta_{\rm{A}}$). Specifically, for each random iteration, the best-fit SEDs from the perturbed data are re-fit, and 1$\sigma$ uncertainties (16$^{\rm{th}}$ and 84$^{\rm{th}}$ percentiles) for the parameters are estimated from the distributions of the fitted, perturbed quantities. 

\begin{table}
    \centering
    \begin{tabular}{rrr}
        \multicolumn{3}{c}{GOODS-S 43114 \& GOODS-S 43683 Physical Properties} \\
        \hline\hline
        Physical Property & GOODS-S 43114 & GOODS-S 43683 \\
        (1) & (2) & (3) \\
        \hline
 $H_{\rm{AB}}$ & 19.843 $\pm$ 0.003 & 21.479 $\pm$ 0.011 \\
  \\
 log$_{10}(M_{\ast}/M_{\odot}$) & 11.64$^{<+0.01}_{<-0.01}$ & 11.04$^{<+0.01}_{<-0.01}$ \\
  \\
 log$_{10}$($t/\tau$) & 0.96$^{<+0.01}_{<-0.01}$ & 0.88$^{+0.01}_{<-0.01}$ \\
   \\
 log$_{10}$($\tau$/yr) & 8.00$^{<+0.10}_{<-0.10}$ & 7.80$^{<+0.10}_{<-0.10}$ \\
   \\
 log$_{10}$(SFR(SED)) (M$_{\odot}$ yr$^{-1}$) & 0.81$^{<+0.01}_{<-0.01}$ & 0.97$^{+0.01}_{-0.06}$ \\
   \\
 log$_{10}$(sSFR(SED)) (yr$^{-1}$) & $-10.83^{<+0.01}_{<-0.01}$ & $-10.07^{<+0.01}_{-0.06}$ \\
   \\
 $A_{\rm{V}}$ & 0.46$^{<+0.01}_{<-0.01}$ & 1.24$^{+0.01}_{-0.02}$ \\
   \\
 U$-$V & 1.50$^{<+0.001}_{<-0.001}$ & 1.56$^{+0.005}_{-0.005}$ \\
   \\
 V$-$J & 0.82$^{<+0.001}_{<-0.001}$ & 1.06$^{+0.007}_{-0.001}$ \\ 
   \\
 D$_{\rm{n}}$4000 & 1.23$^{<+0.001}_{<-0.001}$ & 1.16$^{+0.004}_{-0.001}$ \\
   \\
 H$\delta_{\rm{A}}$ (\AA) & 7.81$^{<+0.01}_{<-0.01}$ & 10.17$^{+0.06}_{-0.01}$ \\
   \\
 $\sigma_{\rm{e}}$ (km s$^{-1}$) & 166 $\pm$ 21 & N/A \\
   \\
 log$_{10}(M_{\rm{dyn}}/M_{\odot})$ & 11.16 $\pm$ 0.11 & N/A \\
   \\
 $R_{\rm{e}}$ (arcseconds) & 0.366 $\pm$ 0.002 & N/A \\
   \\
 S\'ersic Index & 5.26 $\pm$ 0.06 & N/A \\
   \\
 Axis Ratio & 0.865 $\pm$ 0.004 & N/A \\
        \hline
    \end{tabular}
    \caption{
  Col. (1): Physical property of the galaxies in the sample.
  Col. (2): Property value for GOODS-S 43114.
  Col. (3): Property value for GOODS-S 43683. ``N/A'' is listed for $\sigma_{\rm{e}}$ and log$_{10}(M_{\rm{dyn}}$) because we do not have MOSFIRE spectra for this galaxy. Additionally, ``N/A'' is listed for the S\'ersic index, axis ratio, and $R_{\rm{e}}$ due to poor fits. As discussed, the negligible formal uncertainties in the stellar population properties reported here (i.e., $M_*$, $t$, $\tau$, SFR(SED), sSFR(SED), $A_V$, U$-$V, V$-$J, and D$_n$4000) do not reflect the true systematic uncertainties in such quantities.}
    \label{tab:fast_and_spec_property_values}
\end{table}

As discussed in Section \ref{subsec:merger_sed_fitting}, we obtain bootstrap uncertainties on $\sigma_{\rm{e}}$ by perturbing the MOSFIRE $J$-band spectrum within its error spectrum and randomly changing the polynomial degree of fit. The uncertainty on M$_{\rm{dyn}}$ was estimated by propagating the uncertainties on $\sigma_{\rm{e}}$, $R_{\rm{e}}$, $n$, and $L_{\rm{Ser}}$ through Equations \ref{eqn:virial_ml_ratio} and \ref{eqn:dynamical_mass}. 
Finally, the derivation of the uncertainties for the morphological properties adopted from the \citet{van14} catalog (i.e., $R_{\rm{e}}$, $n$, and $q$) is described in \citet{van12}.

Note that many of the reported uncertainties are very small for properties estimated from the SED fitting, which results from both galaxies having both a very high number (39) of photometry points and small photometric error bars. 
In cases where we formally found zero uncertainty, we report upper limit in the magnitude of uncertainties based on either the smallest detected finite uncertainty for GOODS-S 43683 (e.g., U$-$V and V$-$J) or the parameter grid spacing (e.g., $A_{\rm{V}}$ and $\tau$). 
However, it is essential to note that the quoted errors do not reflect systematic uncertainties due to various assumptions inherent to SED fitting models (e.g., stellar initial mass function, treatment of late stages of stellar evolution, dust attenuation curve; \citealt{kri16}). 
\citealt{muz09} report that these systematic uncertainties are typically a few tenths of a dex, but the exact estimate varies for each measured property (e.g., $M_{\ast}$, $t$, etc.).

GOODS-S 43114 is found to be more massive and less dusty with a slightly older stellar population, lower SFR(SED), sSFR(SED), H$\delta_{\rm{A}}$, larger D$_{\rm{n}}$4000, and bluer UVJ colors compared to GOODS-S 43683.  
Note that the observed difference in sSFR(SED) can be mostly attributed to the difference in $M_{\ast}$. 
According to standard classification \citep[e.g.,][]{cox08}, the close stellar mass ratio between GOODS-S 43114 and GOODS-S 43683 implies that this system is a ``major merger.'' We note that \citet{van10} described the merger between GOODS-S 43114 and GOODS-S 43683 as a ``minor merger," based
on an approximate comparison of $H$-band brightnesses. However, based on both our updated comparison of $H$-band brightesses and stellar masses from actual SED fits to both galaxies (indeed, \citealt{van10} only modeled the stellar population of GOODS-S 43114), we find a significantly closer mass ratio between the two galaxies.

Recall from Section \ref{subsec:merger_sed_fitting} that we fit a combination of the photometry and MOSFIRE spectra of GOODS-S 43114, while we only fit the photometry of GOODS-S 43683. The values presented in Table \ref{tab:fast_and_spec_property_values} do not change significantly if we fit GOODS-S 43114 using only the photometry, applying the same methodology as for GOODS-S 43683. In addition, the conclusions we derive from these results (see Section \ref{sec:merger_discussion}) are not dependent on whether we fit both the photometry and MOSFIRE spectra of GOODS-S 43114, or the photometry alone.

Based on rest-frame UVJ colors, \citet{zic18} group galaxies into five bins: quiescent (i), post-starburst (ii), dusty galaxies with lower sSFRs (iii), dusty star-forming (iv), and non-dusty star-forming (v). These bins were created based on known correlations between galaxy properties and UVJ colors. For star-forming galaxies, $A_{\rm{V}}$ increases linearly with increasing U$-$V and V$-$J. sSFR(SED) decreases along a direction perpendicular to this $A_{\rm{V}}$ sequence (i.e., U$-$V increases and V$-$J decreases as sSFR(SED) decreases; e.g., \citealt{yan16}). These relationships motivate the creation of  bins (iii), (iv), and (v). Once a galaxy has stopped forming stars its colors will evolve towards the quiescent box. As galaxies age away from their star forming epochs, they move along the quiescent sequence in the UVJ diagram (i.e., increasing in U$-$V and V$-$J color). Therefore, bins (i) and (ii) comprise an age gradient for galaxies no longer forming stars (e.g., \citealt{whi12, bel19}). 

The left panel of Figure \ref{fig:zick2018_plots} gives the UVJ diagram with these galaxy bins from \citet{zic18} identified. The region occupied by quiescent galaxies following the definition from \citet{wil09} is also included. 
It can be seen that GOODS-S 43114 falls well within the post-starburst bin (i.e., bin (ii)), while GOODS-S 43683 falls at the intersection of bins (i), (ii), and (iii) --- reflecting its redder V$-$J color. 
The SFR(SED) and sSFR(SED) values in both galaxies indicate that these systems are not currently forming stars at a rapid rate relative to their past averages. In addition, the characteristic shapes of the SEDs in the right panel of Figure \ref{fig:fast_fit_and_spectra} (specifically, the steep decline in flux density bluewards of $\sim$4000 \AA) suggests that both galaxies are in a post-starburst phase. 

Figure \ref{fig:zick2018_plots} shows GOODS-S 43114 and GOODS-S 43683 on the H$\delta_{\rm{A}}$ vs. D$_{\rm{n}}$4000 diagram. 
Also included are the five datapoints from \citet{zic18} representing stacks of galaxies in each of the five UVJ bins, and tracks of SFH with $\tau$ = 0.1, 0.2, and 1.0 Gyr. 
To create these tracks, we use the code \texttt{python-FSPS}\footnote{https://dfm.io/python-fsps/current/} \citep{joh21} which generates spectra for synthetic stellar populations using the FSPS library \citep{con10}.
For these models, we make similar assumptions to those in our SED fitting (i.e., a \citealt{cha03} stellar initial mass function (IMF), a \citealt{cal00} dust attenuation curve, and delayed-$\tau$ SFHs). 

We find that both GOODS-S 43114 and GOODS-S 43683 fall within the uncertainties of bin (ii) on the H$\delta_{\rm{A}}$ vs. D$_{\rm{n}}$4000 diagram. 
Based on the elevated H$\delta_{\rm{A}}$ value found for both galaxies, it is likely that the spectra are dominated by A-type stars, which implies that star formation shut down in a time frame where the O- and B-type stars no longer exist but the A-type stars still remain. 
In addition, the fact that the D$_{\rm{n}}$4000 feature and $t/\tau$ ratio are very similar for both galaxies indicates that both galaxies have similar stellar population ages, and therefore shut off star formation at roughly the same time. 

This approximate synchronization is confirmed in Figure \ref{fig:sf_histories}, which shows the estimated SFHs for GOODS-S 43114 and GOODS-S 43683 obtained using the FAST best-fit SFR(SED), $t$, and $\tau$ values from Table \ref{tab:fast_and_spec_property_values}. 
Also shown are the SFHs estimated using \texttt{Prospector}\footnote{https://prospect.readthedocs.io/en/latest/} \citep{joh21prospector}, which utilizes ``non-parametric'' SFHs (i.e., models that do not assume a functional form for SFR as a function of time). 
For the \texttt{Prospector} modeling, we adopt the FSPS stellar library \citep{con10} and assume a \citet{cha03} IMF with a 300 M$_{\odot}$ upper limit, a \citet{cal00} dust attenuation curve, and fix the metallicity to solar (0.019). We split the non-parametric SFH into eight time bins, assume constant star formation in each bin, and adopt a built-in continuity prior that weights against sharp variations in SFR between adjacent time bins (see \citealt{tac22} for discussion on how non-parametric models are influenced by the choice in priors used). The first two age bins are fixed at 0-30 Myr and 30-100 Myr, with the remaining bins spaced logarithmically from 100 Myr to the age of the universe at $z=1.89$.
We include an additional prior placing an upper limit of 10 M$_{\odot}$ yr$^{-1}$ on the SFR in the earliest time bins ($z>4$) and also the time bin closest to the epoch of observation, for consistency with the strong Balmer absorption observed in the $J$-band spectrum of GOODS-S 43114. 

While there are differences between the inferred SFHs using FAST and Prospector in detail, in terms of the precise SFH shape and lookback time of peak star formation, we emphasize the qualitative similarities here. 
Both the FAST delayed-$\tau$ and \texttt{Prospector} non-parametric SFHs suggest that the onsets and peaks of the most recent bursts of star formation in these two galaxies are synchronized to within a few hundred Myrs of each other. There is uncertainty between the models on the exact timing of the onset and peak of the burst, as well as peak level of SFR; however, it is clear that the star-forming events in these two galaxies are linked. We note that \citet{van10} model the multi-wavelength SED and rest-optical spectrum of GOODS-S 43114 using simple top-hat SFHs, also finding that star formation recently ceased and must have been significantly higher in the past. In Section~\ref{subsec:merger_state}, we discuss the impact of the galaxy interaction on the star-formation histories of GOODS-S 43114 and GOODS-S 43683.

\subsection{Second Velocity Moment \& Dynamical Mass of GOODS-S 43114}

\begin{figure}
    \includegraphics[width=0.98\linewidth]{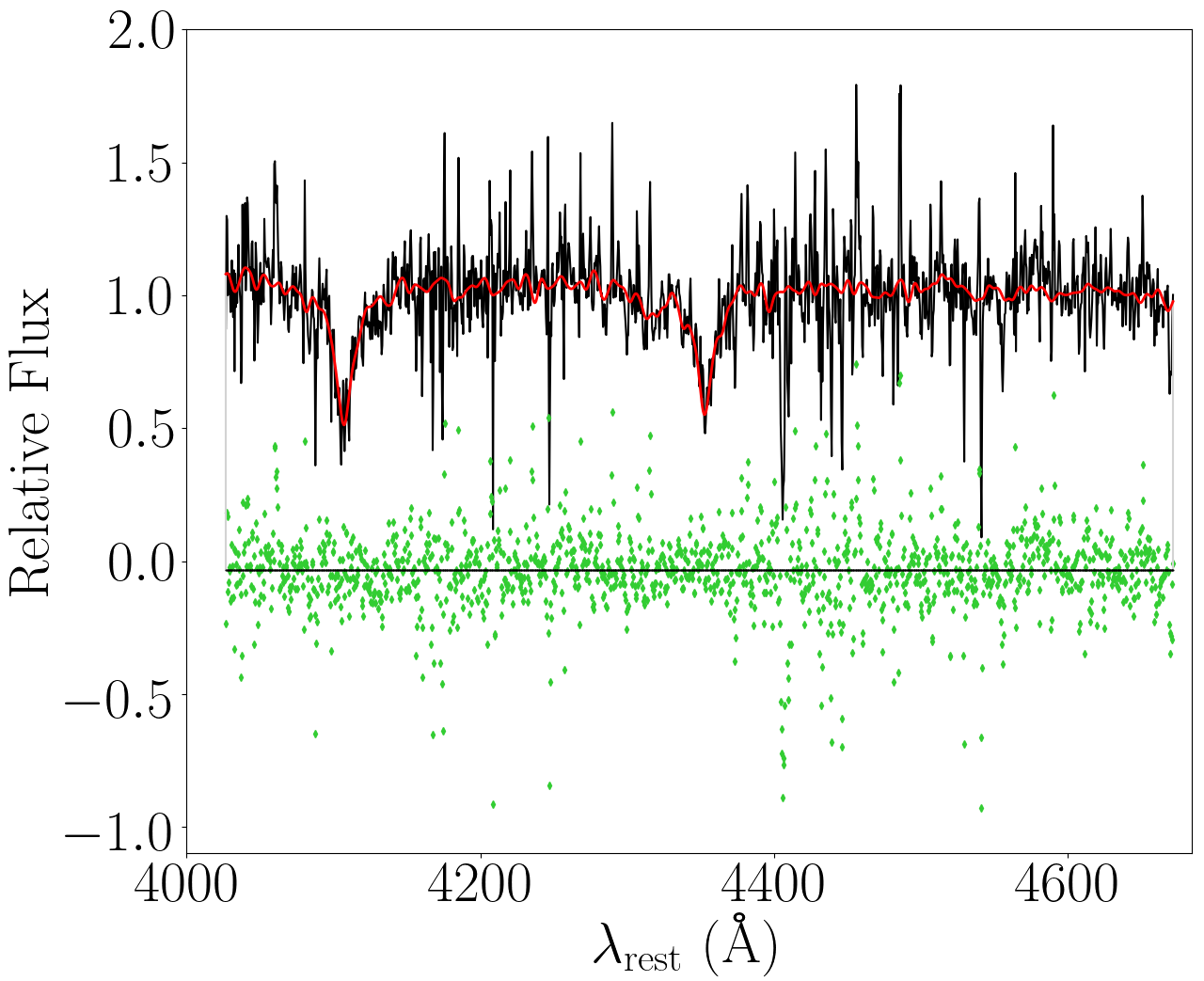}
    \caption{Fit to the MOSFIRE $J$-band spectrum of GOODS-S 43114 using \texttt{ppxf} to estimate $\sigma_{\rm{e}}$. The MOSFIRE spectrum is shown in black while the fit is shown in red. The residuals to the fit are given as green points. The spectrum and fit are centered at a relative flux of 1, while the residuals are centered at a relative flux of 0.}
    \label{fig:ppxf_plot}
\end{figure}

As discussed, the MOSFIRE $J$-band spectrum enables an estimate of the dynamical properties of GOODS-S 43114. Using \texttt{ppxf}, we find $\sigma_{\rm{e}}$ = 166 $\pm$ 21 km s$^{-1}$. Combining equations \ref{eqn:virial_ml_ratio} and \ref{eqn:dynamical_mass} results in log$_{10}(M_{\rm{dyn}}/M_{\odot})$ = 11.16 $\pm$ 0.11 for GOODS-S 43114. Figure \ref{fig:ppxf_plot} shows the fit to the $J$-band spectrum with the residuals of the fit. The dynamical mass estimate is approximately a factor of three lower than the $M_{\ast}$ measurement, which raises concerns about the validity the method used to derive the dynamical mass. We discuss the discrepancy between the mass estimates in Section \ref{subsec:M_dyn_discussion}. Since we do not have MOSFIRE spectra of GOODS-S 43683, we cannot estimate $\sigma_{\rm{e}}$ and subsequently M$_{\rm{dyn}}$ for this galaxy.

\section{Discussion} \label{sec:merger_discussion}

\subsection{State of the Merger} \label{subsec:merger_state}

Comparing observations to galaxy simulations is important for connecting isolated snapshots of galaxies with the larger evolutionary picture. 
Numerical simulations help improve our understanding of the both the merger process and the significance of mergers to galaxy formation (e.g., \citealt{bar91, bar96, di07, cox08, di08, mor15, blu18}). State-of-the-art galaxy formation simulations have the resolution to capture feedback-regulated star formation and multi-phase structures within the interstellar medium (ISM; e.g., FIRE-1 and FIRE-2; \citealt{hop11, hop14, hop18}), and, therefore, how these galaxy processes and components will respond to a merger event. For example, recent merger modeling with FIRE-2 \citep{mor19} probes enhanced star formation and ISM gas content during the merger process. In this work, three key stages in a merger are identified: the first pericentric passage, the second pericentric passage, and coalescence. The time between the first and second pericentric passage is $\sim$2-2.5 Gyr while the time between the second pericentric passage and coalescence is $\lesssim$0.5 Gyr. In addition, \citet{mor19} suggest that SFR spikes at each of these three events with a significant decrease in SFR during the large time gap between the pericentric passages. The SFR stays elevated in the short time between second pericentric passage and coalescence. 

One important caveat to note is that \citet{mor19} simulated mergers for local galaxies. Additionally, the masses of the galaxies in the \citet{mor19} fiducial model are an order of magnitude lower than those of the galaxy pair in this study. Such differences (in addition to the detailed orbital configurations of the simulated and observed mergers) will likely result in different absolute timescales for the merger and the corresponding rise and fall in the SFRs of the merging galaxies. While the precise timescales between pericentric passages and the magnitude of the SFR spikes resulting from these events may be different for the high-mass $z\sim2$ galaxy pair investigated in this study, we assume that the basic pattern of galaxies having an elevation in SFR at each pericentric passage with a decrease in-between still applies \citep[see also, e.g.,][]{cox08,hop08}. Therefore, we only make a qualitative comparison with the \citet{mor19} fiducial model, focusing generally on the phase of the merging process in which GOODS-S 43114 and GOODS-S 43683 are observed based on their galaxy properties and recent SFHs at the epoch of observation.

For the merger analyzed here, the low SFR(SED) and sSFR(SED) values of GOODS-S 43114 and GOODS-S 43683 compared to their past averages indicate that these galaxies are not observed during one of the merger phases corresponding to elevated star formation. Additionally, Figure \ref{fig:sf_histories} indicates that both galaxies had much higher SFRs in the past. The tidal tail features exhibited by both galaxies suggest that this system has undergone at least one pericentric passage. Due to the depressed SFR at the date of observation, it is likely that the merger is observed between the first and second pericentric passages. Furthermore, the the SFHs shown in Figure \ref{fig:sf_histories} show that the recent burst of star formation began $\sim$0.5-1 Gyr (depending on the model) before the data of observation, suggesting that the first pericentric passage occurred around that time. 

The elevated H$\delta_{\rm{A}}$ values and SFHs shown in Figure \ref{fig:sf_histories} for GOODS-S 43114 and GOODS-S 43683 suggest that both galaxies have recently shut off their star formation. The main sequence lifespan of O-, B- and A-type stars are $\sim$0.01, $\sim$0.1 and $\sim$1 Gyr, respectively. A-type stars will dominate the stellar spectrum during their main sequence lifetime after the main sequence lifetimes of O- and B-type stars have passed. For both the FAST and \texttt{Prospector} models, the time since peak star formation is longer than the O- and B-star lifetime, corresponding to a phase when the light from A stars will contribute significantly to the rest-frame optical spectrum. 

Combining all of the information gained from the SED fitting with inferences from the models indicates that these galaxies recently underwent their first pericentric passage and are moving away from each other. Detailed velocity mapping would be needed to confirm this theory, and such data currently do not exist for this merger. 

As discussed earlier in this section, it is difficult to make direct comparisons with simulated mergers as a variety of parameters can alter the timescales and elevation/depression of star formation. Aside from galaxy mass and redshift, additional caveats are differences in other merger parameters relative to the \citet{mor19} fiducial model. For example, different orientations and levels of gas content in the galaxies can affect the time between pericentric passages and the elevation in SFR. \citet{fen17} show that mergers with 10\% gas fractions characteristic of $z\sim0$ galaxies undergo a larger increase in SFR compared to mergers of more gas-rich (60\% gas fraction) systems characteristic of  $z\sim2$ galaxies. Along the same lines, \citet{scu15} find an anti-correlation between SFR enhancement during the merger and the initial gas fraction. This dependence of SFR enhancement during the merger on gas content can be attributed to a correlation between ISM turbulence and central gas inflow during the merger. In the \citet{fen17} models, both the local-analog gas-poor mergers and $z\sim2$-analog gas-rich mergers display similar peak gas inflow rates that fuel star formation. The difference in SFR enhancement between the two scenarios is therefore due to the fact that the $z\sim2$-analog gas-rich galaxies begin the merging process with much higher gas gas reservoirs.

\subsection{Implications of the Apparent Mass Discrepancy} \label{subsec:M_dyn_discussion}

The large discrepancy between $M_{\ast}$ and $M_{\rm{dyn}}$ for GOODS-S 43114, in particular with $M_{\rm{dyn}} < M_{\ast}$, raises concerns about the accuracy of these measurements. As stated in Section \ref{subsec:merger_sed_fitting}, our methodology for estimating $M_{\ast}$ is robust. 
In Table \ref{tab:fast_and_spec_property_values}, we list the $M_{\ast}$ estimated from the delayed-$\tau$ model. Note that we experimented with other parametric SFHs (constant, exponentially rising, and exponentially falling), as well as a \texttt{Prospector} non-parametric model, and find similar $M_{\ast}$ to that inferred from the delayed-$\tau$ model (i.e., the choice of a different SFH does not resolve the tension between the inferred stellar and dynamical masses. 
The SED is fit assuming a realistic delayed-$\tau$ SFH and we use 39 photometry points to constrain the shape of the SED. Additionally, we not only find consistent results using non-parametric modeling with the \texttt{Prospector} code, but also our estimate of $M_{\ast}$ agrees within uncertainties with past studies \citep{van10}. 

Therefore, it is likely that the $M_{\rm{dyn}}$ measurement is significantly underestimated. There is a straightforward explanation for how such an underestimate can arise.
The axis ratio of the galaxy obtained from the best fit S\'ersic model is 0.865 $\pm$ 0.004. In one plausible scenario, this large axis ratio would result if GOODS-S 43114 is a roughly face-on disk. In such a geometry, the estimate of the second velocity moment (and subsequent $M_{\rm{dyn}}$ estimate) is biased low relative to the true value. 

For a more in-depth estimate of $M_{\rm{dyn}}$, we compute the dynamical mass of the galaxy by constructing a cylindrically-aligned Jeans Anisotropic Model (JAM$_{\rm{cyl}}$; \citealt{cap08}) of its S\'ersic approximation. The advantage of this approach with respect to using a S\'ersic-dependent virial equation is that JAM allows one to explore the effect of inclination as well as modelling the effect of the PSF and the kinematics extraction aperture. We first approximated a S\'ersic profile with the index $n = 5.26$ given by \citet{van14} with 30 Gaussians using the Multi-Gaussian Expansion (MGE) parametrization and the mge\_fit\_1d procedure\footnote{We used v5.0 of the MgeFit Python package from https://pypi.org/project/mgefit/} of \citet{cap02}. Then we used the fitted MGE to compute the velocity second moment using the jam\_axi\_proj procedure in the Jeans Anisotropic Modelling (JAM) software package\footnote{We used v6.3 of the JamPy Python package from https://pypi.org/project/jampy/} of \citet{cap08}. We also included a central black hole with a mass of 0.2\% of the total mass of the S\'ersic model. This value is half of the expected value in the local Universe \citep{kor13} and has minimal effect on the result. We adopted a unitary mass for the S\'ersic model, and assumed both that mass follows light and a typical anisotropy $\beta = 0.2$ (Figure 9 of \citealt{cap16}) to compute the PSF-convolved JAM model predictions for the line-of-sight velocity second moments $\overline{v^{2}_{\rm{los},j}}$ on a dense grid of values covering the observed rectangular aperture over which the MOSFIRE spectrum was extracted. The model is placed at the angular-size distance ($D_{\rm{A}}$) for the adopted standard cosmology. We then co-added these luminosity-weighted values inside the aperture as $\overline{v^{2}_{\rm{los}}} = \Sigma_{j}I_{j}\overline{v^{2}_{\rm{los},j}}/\Sigma_{j}I_{j}$, where $I_{\rm{j}}$ is the PSF-convolved galaxy surface brightness at the $j$-th grid location. 

Given the adopted unitary mass of the JAM model and the proportionality between masses and squared velocities, the galaxy dynamical mass can be computed as $M_{\rm{JAM}} = V^{2}_{\rm{rms}}/v^{2}_{\rm{los}}$, where $V_{\rm{rms}}$ is the observed velocity dispersion fitted by \texttt{pPXF} from the MOSFIRE spectrum, which approximates the luminosity-weighted second velocity moments inside the aperture (see discussion after equation 29 of \citealt{cap13}). The fitted dynamical mass physically represents the stellar mass of the galaxy, under the assumption that the stellar mass follows the light and the $M/L$ has the same value as its average $(M/L)(r < R_{\rm{ap}})$ inside the region $R_{\rm{ap}} \lesssim 1.35R_{\rm{e}}$ covered by the MOSFIRE kinematics:
\begin{equation}
    M_{\rm{JAM}} \approx L_{\rm{Ser}} \times (M/L)(r < R_{\rm{ap}})
\end{equation}
If the galaxy contains dark matter, then $M_{\rm{JAM}}$ should be decreased by a fraction $1 - f_{\rm{DM}}(r = R_{\rm{ap}})$ (equation 23 of \citealt{cap13}) to obtain an estimate of the galaxy stellar mass.

As extreme cases, we obtained log$_{10}(M_{\rm{JAM}}/M_{\odot}) = 11.19 \pm 0.10$ when assuming the galaxy has the intrinsic axial ratio $q_{\rm{intr}} = 0.2$ of a flat stellar disk, in which case it is seen at an inclination $i = 31^{\circ}$, and log$_{10}(M_{\rm{JAM}}/M_{\odot}) = 11.06 \pm 0.10$ when assuming the galaxy is edge-on ($i = 90^{\circ}$) and has the same $q_{\rm{intr}} = 0.87$ as the observed isophotes. We obtained insignificant differences in the two extreme assumptions that the MOSFIRE slit was aligned with either the major or minor projected axis of the galaxy. As a final test, we disregard the observed axial ratio and assume as the most extreme case that the galaxy is perfectly face-on, still with $q_{\rm{intr}} = 0.2$ and ansiotropy $\beta = 0.2$. This unrealistic scenario results in log$_{10}(M_{\rm{JAM}}/M_{\odot}) = 11.44$, still less than the measured $M_{\ast}$ from FAST. Therefore, the discrepancy between $M_{\ast}$ and $M_{\rm{dyn}}$ cannot be resolved by simply taking into consideration the observed geometry of the galaxy. Further investigation into solving this mass inconsistency is outside the scope of this paper, but will likely need to incorporate the effect of the merger on the internal dynamics of GOODS-S 43114. 

Better consistency between dynamical and stellar mass measurements has been found in other studies of high-redshift quiescent galaxies.
For example, \citet{van13} find dynamical masses $\sim 15$\% larger than stellar masses for a sample of 5 massive ($\log_{10}(M_{\ast}/M_{\odot}) > 11$) quiescent $z\sim 2$ galaxies. Additional studies of high-redshift, massive, quiescent galaxies find that $M_{\ast}$ and $M_{\rm{dyn}}$ are roughly consistent with each other (e.g., \citealt{bel17, tan19, esd21}). However, the majority of galaxies in these earlier works have structural properties consistent with a spheroidal geometry. 

GOODS-S 43114 is not unique for appearing disk-like at $z\sim2$. 
For example, \citet{new18a, new18b} identify four fast-rotating ($v/\sigma\sim 2$), disk-dominated quiescent galaxies at $z\sim2$ with comparable stellar masses to those presented here \citep[see also][]{tof12,tof17}. \citeauthor{new18b} and \citeauthor{tof17} argue that merger interactions will cause these massive galaxies evolve into the slow rotator ETGs that dominate above $M_{\ast}\gtrsim2\times10^{11}$ M$_{\odot}$ in the nearby Universe (see review by \citealt{cap16}), given their already-extreme masses.

GOODS-S 43114 is undergoing a major merger interaction with GOODS-S 43683, which will likely have a significant impact on its structural properties. Its subsequent star-formation history to the present day will depend on the details of the merger interaction with GOODS-S 43683 as well as the cool gas content in both galaxies. Accordingly, both GOODS-S 43114 and GOODS-S 43683 comprise compelling targets for observations of molecular gas (e.g., CO), in order to understand the fuel available for subsequent star formation.

\subsection{Constraints from Far-IR and Radio Observations}

Up until this point, the longest wavelength data analyzed here has been the IRAC Channel 4 band at 8 $\mu$m ($\sim2.75 \ \mu$m in the rest frame). In this section, we fold longer wavelength data into the discussion, specifically the 24 $\mu$m band from the Multiband Imaging Photometer for Spitzer (MIPS; \citealt{rie04}) instrument on the \textit{Spitzer Space Telescope} \citep{wer04}; the 70, 100, and 160 $\mu$m bands from the Photodetector Array Camera and Spectrometer (PACS; \citealt{pog10}) and the 250, 350, and 500 $\mu$m bands from the Spectral and Photometric Imaging REceiver (SPIRE; \citealt{gri10}) instruments on the \textit{Herschel Space Observatory}; and the 1.4 GHz band from the \textit{Very Large Array} (VLA; \citealt{per11,mil13,big11}, Smail et al., private communication). We obtained MIPS data from \citet{gia04}, and PACS and SPIRE observations from \citet{elb11}. These longer-wavelength data points more directly trace the dust content of galaxies and can reveal star formation and AGN activity that is severely obscured by dust at rest-frame UV through rest-frame near-IR wavelengths. For example, \citet{sma99} show in a sample of $z=0.4$ cluster galaxies that some galaxies previously classified as post-starbursts from rest-frame optical data alone contain radio-bright components revealing previously unaccounted-for, obscured star formation.

\begin{figure}
    \includegraphics[width=0.98\linewidth]{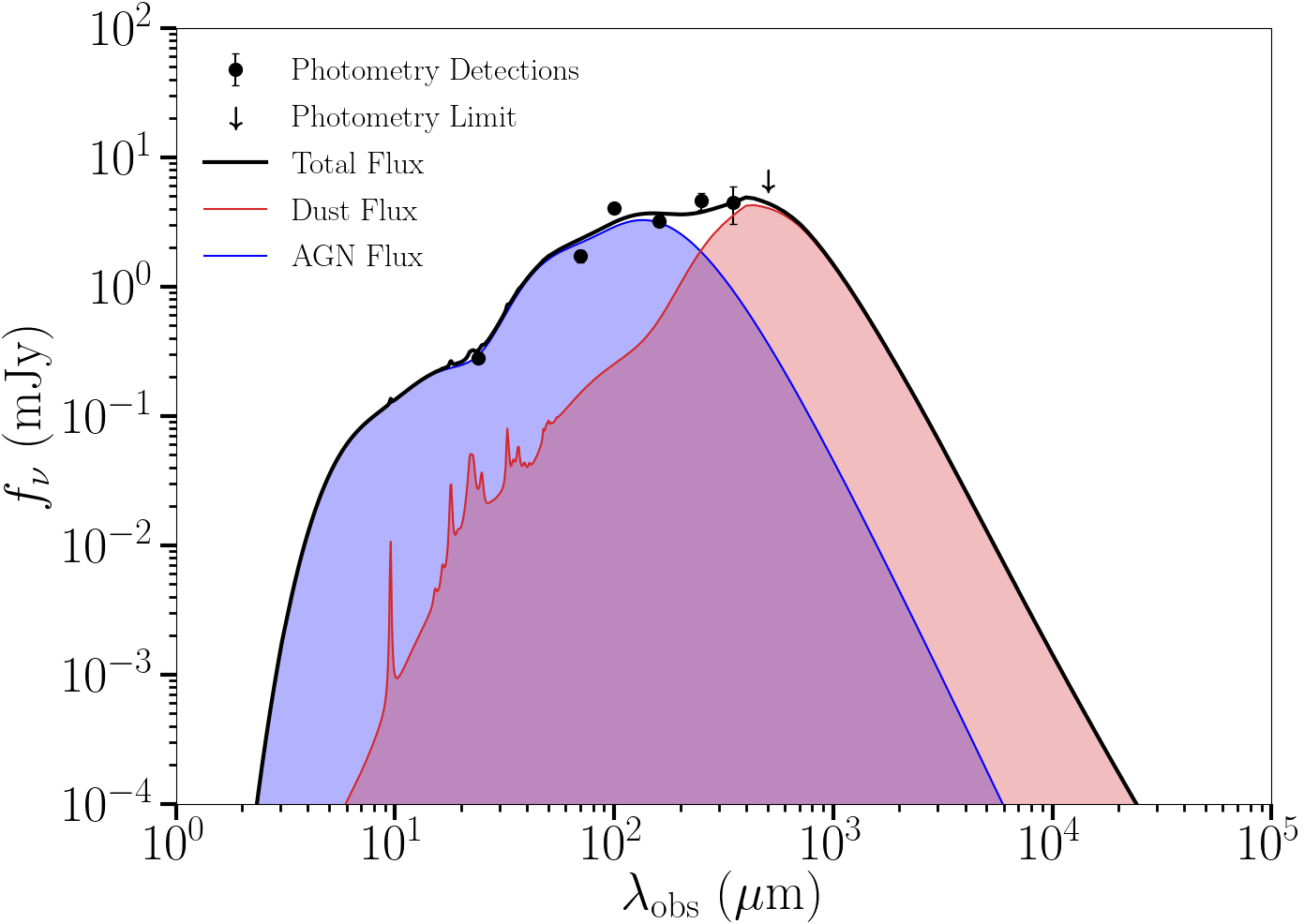}
    \caption{Fit to the \textit{Spitzer}/MIPS 24 $\mu$m, \textit{Herschel}/PACS 70, 100, and 160 $\mu$m, and \textit{Herschel}/SPIRE 250, 350, and 500 $\mu$m far-IR flux measurements of the GOODS-S 43114+GOODS-S 43683 merging system. The galaxy pair is unresolved at these wavelengths, so the MIPS, PACS, and SPIRE measurements represent the summed contributions from both galaxies. The SED-fitting code, \texttt{Stardust}, is used to constrain the longer wavelength portion of the SED for the galaxy pair and estimate the AGN (blue) and dust (red) components of the total flux (black). The photometric data points and associated uncertainties are shown with black data points, except for the 500 $\mu$m SPIRE band which is given as an arrow because it is a limit.}
    \label{fig:farIR_plot}
\end{figure}

Long wavelength observations of the merger system in fact reveal detections at mid- and far-IR wavelengths, indicating the presence of dust emission. However, the angular resolution of MIPS, PACS, and SPIRE is not sufficient to secure spatially-distinct of GOODS-S 43114 and GOODS-S 43683 (i.e., the two galaxies are blended together). 
Therefore, in this analysis, we attempt to constrain the total dust-obscured SFR and AGN contribution associated with the combined 43114+43683 system (as opposed to estimating the dust-obscured SFR and AGN contribution in each galaxy individually). We can use this combined SFR estimate to test if both galaxies are found in a post-starburst phase, even when long-wavelength data is taken into account. The methodology adopted here avoids uncertain assumptions about the distribution of mid-IR and far-IR flux between the two galaxies. 
We use the code \texttt{Stardust} \citep{kok21} to model the observed mid- and far-IR SED of the 43114+43683 pair. 
The \texttt{Stardust} code is ideal for this analysis, given the known AGN detections of GOODS-S 43114 and GOODS-S 43683, because \texttt{Stardust} simultaneously accounts for the multi-wavelength contributions from stars, AGN, and infrared dust emission. 
Furthermore, \texttt{Stardust} does not require energy balance, thus allowing for additional star formation that is completely obscured at rest-frame UV through near-IR wavelengths \citep{kok21}. 
Note that we use the \citet{mul11} templates to fit the AGN component, the \citet{dra07, dra14} templates to fit the IR dust, and utilize \citet{jin18} for the deblending technique in the extraction of far-IR photometry. 

Figure \ref{fig:farIR_plot} shows the mid- and far-IR SED for the 43114+43683 merger system, along with the best-fit multi-component \texttt{Stardust} model. The blue shaded region identifies the portion of the SED attributed to light from the AGN ($L_{\rm IR,AGN}=(1.14\pm0.02)\times 10^{12}$ L$_{\odot}$) while the red shaded region identifies the portion attributed to dust emission that has been heated by the stellar population ($L_{\rm IR,SF}=(3.1\pm0.6)\times 10^{11}$ L$_{\odot}$, with an associated $M_{\rm dust}=(2.8\pm0.5)\times 10^9$ M$_{\odot}$). The black curve represents the total AGN+dust SED. We find that the resultant SFR associated with the non-AGN portion of the IR SED is 31$\pm$6 M$_{\odot}$ yr$^{-1}$, which is not significantly elevated compared with our FAST estimate ($\sim$16 M$_{\odot}$ yr$^{-1}$). 
We note that the inferred 1.4 GHz flux density of $30 \pm 7$ $\mu$Jy (while not included in the \texttt{Stardust} fitting) is entirely consistent with the best-fit AGN+dust model fit to the combined mid- and far-IR SED following the $L_{\rm{IR}}$ to radio correlation from \citet{del17}. 

The SFR estimate inferred from the fit to the long-wavelength data, while larger than that inferred from the SED at \textit{Spitzer}/IRAC and shorter wavelengths, is also significantly lower than the peak SFRs inferred several hundred Myr in the past from both FAST and \texttt{Prospector} (see Figure \ref{fig:sf_histories}). 
Additionally, the higher SFR value from \texttt{Stardust} could be due to a time lag, as the SFR estimated from long wavelengths is less responsive to recent changes in the SFR compared to the rest-optical regime. If the SFR of the system is actively decreasing, which is expected between the first and second pericentric passages (see \citet{mor19}), the SFR estimated from dust could easily be larger than the SFR inferred from the rest-optical regime. 
Therefore, we conclude that the inclusion of the mid- and far-IR data does not change the principal conclusions of this work: both galaxies still appear to be in a post-starburst phase, which is consistent with multiple pieces of evidence (i.e., strong Balmer absorption lines and SFHs fit to rest-frame UV through rest-frame near-IR SEDs) presented in this study. 

As we look ahead, JWST and/or ALMA will be needed to obtain distinct observations of GOODS-S 43114 and GOODS-S 43683 (i.e., not blended together) at long wavelengths. Such data will enable more robust constraints on the dust content, AGN activity, and obscured star formation, individually, in each galaxy.
We also note that \citet{van10} find evidence for faint spiral arms in the residuals of the F160W image of GOODS-S 43114, once the dominant compact core is subtracted off. The spatially-resolved  {\it HST}/WFC3 grism spectrum of GOODS-S 43114 may suggest a faint level of residual star formation associated with these spiral arms.

\section{Summary} \label{sec:merger_summary}

In this study we analyze high-quality Keck/MOSFIRE spectra and multi-wavelength photometry of a merger at $z=1.89$ between two galaxies that host AGN: GOODS-S 43114 and GOODS-S 43683. 
Stellar Balmer absorption lines (H$\gamma$ and H$\delta$) are detected in the MOSFIRE $J$-band of GOODS-S 43114. 
Combining the high S/N spectra and broadband SED-fitting with insights from realistic galaxy merger simulations enables us to obtain and study the physical properties and SFHs of both galaxies in the merger pair as well as the current phase of the merger process. 

The main results are as follows:
\begin{enumerate}
    \item Both merging galaxies have a close $M_{\ast}$ ratio, indicating that this is a major merger. In addition, both galaxies have recently shut off their star formation as shown by their SFR(SED) and strong Balmer absorption lines. 
    \item The depressed star formation rates combined with the visible tidal tails imply that the galaxies have undergone their first pericentric passage. Modeling the SFHs suggests that the pericentric passage most likely happened $\sim$0.5-1 Gyr ago (approximately around the time of peak star formation of the system).
    \item GOODS-S 43114 has a low inclination indicating that it is roughly a face-on disk based on its morphology and smaller than expected second velocity moment estimate. 
    More robust modeling that takes into account the inclination and geometry of a face-on disk does not resolve the discrepancy between the dynamical and stellar masses. More work is needed to address this mass inconsistency. Galaxies that are massive ($\log_{10}(M_{\ast}/M_{\odot}) > 11$), disk-shaped, rotating, and quiescent at $z\sim2$ are likely the progenitors of the most massive, spheroidal and slow-rotating elliptical galaxies present in the local universe. However, they must undergo significant structural evolution over the intervening $\sim 10$~Gyr.
    \item Fitting the spatially-unresolved mid- and far-IR SED of the combined GOODS-S 43114 and GOODS-S 43683 system yields consistent conclusions to those obtained by fitting the rest-frame UV through near-IR data of the individual merging galaxies. Specifically, we still find that the GOODS-S 43114 and GOODS-S 43683 system is in a post-starburst phase with ongoing AGN activity, and significantly lower star formation at the epoch of observation compared to the past peak. One limitation of the longer wavelength data is that the 2'' separation between GOODS-S 43114 and GOODS-S 43683 is smaller than the angular resolution of \textit{Spitzer}/MIPS, and \textit{Herschel}/PACS and SPIRE. JWST and/or ALMA observations are needed to study these galaxies in this long-wavelength regime individually.
\end{enumerate}

Understanding the rapid onset and shut-down of star formation during mergers, especially at $z\sim2$ (i.e., the peak epoch of star formation), is important for furthering our knowledge of how mergers affect the galactic baryon cycle. Comprehensive studies of individual mergers like the GOODS-S 43114/GOODS-S 43683 system can help further constrain simulations leading to a more accurate galaxy evolution models. For the merger pair analyzed in this work, space-based observations with the James Webb Space Telescope will enable us to observe the full range of rest-optical emission-lines (e.g., H$\alpha$, H$\beta$, [O~\textsc{III}]$\lambda\lambda$4960,5008, and [O~\textsc{II}]$\lambda\lambda$3727,3730) that are currently hidden outside the windows of atmospheric transmission. 
Measurement of these emission-lines will greatly increase our knowledge of the merger, as they contain a wealth of information on AGN activity and gas outflows, as well as the physical properties of the ionized ISM.

\section*{Acknowledgements}
We acknowledge support from NSF AAG grants AST1312780, 1312547, 1312764, 1313171, 2009313, 2009085, 2009278, grant AR-13907 from the Space Telescope Science Institute, and grant NNX16AF54G from the NASA ADAP program. We also acknowledge a NASA contract supporting the ``WFIRST Extragalactic Potential Observations (EXPO) Science Investigation Team'' (15-WFIRST15-0004), administered by GSFC. Support for this work was also provided through the NASA Hubble Fellowship grant \#HST-HF2-51469.001-A awarded by the Space Telescope Science Institute, which is operated by the Association of Universities for Research in Astronomy, Incorporated, under NASA contract NAS5-26555.
We thank the 3D-\textit{HST} collaboration, who provided spectroscopic and photometric catalogs used to select MOSDEF targets and to derive stellar population parameters. 
We acknowledge useful conversations with Ian Smail that benefited this work.
This research made use of Astropy,\footnote{http://www.astropy.org} a community-developed core Python package for Astronomy \citep{ast13, ast18}. Finally, we wish to extend special thanks to those of Hawaiian ancestry on whose sacred mountain we are privileged to be guests. 

\section*{Data Availability}
The data underlying this article will be shared on reasonable request to the corresponding author.

\vspace{5mm}
\textit{Facilities}: Keck/MOSFIRE, \textit{SDSS}

\textit{Software}: Astropy \citep{ast13, ast18}, EAZY \citep{bra08}, FAST \citep{kri09}, IPython \citep{per07}, JAMPy \citep{cap08}, Matplotlib \citep{hun07}, MGEfit \citep{cap02}, NumPy \citep{van11, har20}, Pandas \citep{mckinney-proc-scipy-2010, pandas20}, pPXF \citep{cap04, cap17}, Prospector \citep{joh21prospector}, python-FSPS \citep{joh21}, SciPy \citep{oli07, mil11, vir20}, Starburst \citep{kok21}

\bibliographystyle{mnras}
\bibliography{references} 

\begin{thebibliography}{}
\makeatletter
\relax
\def\mn@urlcharsother{\let\do\@makeother \do\$\do\&\do\#\do\^\do\_\do\%\do\~}
\def\mn@doi{\begingroup\mn@urlcharsother \@ifnextchar [ {\mn@doi@}
  {\mn@doi@[]}}
\def\mn@doi@[#1]#2{\def\@tempa{#1}\ifx\@tempa\@empty \href
  {http://dx.doi.org/#2} {doi:#2}\else \href {http://dx.doi.org/#2} {#1}\fi
  \endgroup}
\def\mn@eprint#1#2{\mn@eprint@#1:#2::\@nil}
\def\mn@eprint@arXiv#1{\href {http://arxiv.org/abs/#1} {{\tt arXiv:#1}}}
\def\mn@eprint@dblp#1{\href {http://dblp.uni-trier.de/rec/bibtex/#1.xml}
  {dblp:#1}}
\def\mn@eprint@#1:#2:#3:#4\@nil{\def\@tempa {#1}\def\@tempb {#2}\def\@tempc
  {#3}\ifx \@tempc \@empty \let \@tempc \@tempb \let \@tempb \@tempa \fi \ifx
  \@tempb \@empty \def\@tempb {arXiv}\fi \@ifundefined
  {mn@eprint@\@tempb}{\@tempb:\@tempc}{\expandafter \expandafter \csname
  mn@eprint@\@tempb\endcsname \expandafter{\@tempc}}}

\bibitem[\protect\citeauthoryear{{Abazajian} et~al.,}{{Abazajian}
  et~al.}{2009}]{aba09}
{Abazajian} K.~N.,  et~al., 2009, \mn@doi [\apjs]
  {10.1088/0067-0049/182/2/543}, \href
  {https://ui.adsabs.harvard.edu/abs/2009ApJS..182..543A} {182, 543}

\bibitem[\protect\citeauthoryear{{Asplund}, {Grevesse}, {Sauval}  \&
  {Scott}}{{Asplund} et~al.}{2009}]{asp09}
{Asplund} M.,  {Grevesse} N.,  {Sauval} A.~J.,   {Scott} P.,  2009, \mn@doi
  [\araa] {10.1146/annurev.astro.46.060407.145222}, \href
  {https://ui.adsabs.harvard.edu/abs/2009ARA&A..47..481A} {47, 481}

\bibitem[\protect\citeauthoryear{{Astropy Collaboration} et~al.,}{{Astropy
  Collaboration} et~al.}{2013}]{ast13}
{Astropy Collaboration} et~al., 2013, \mn@doi [\aap]
  {10.1051/0004-6361/201322068}, \href
  {https://ui.adsabs.harvard.edu/abs/2013A&A...558A..33A} {558, A33}

\bibitem[\protect\citeauthoryear{{Astropy Collaboration} et~al.,}{{Astropy
  Collaboration} et~al.}{2018}]{ast18}
{Astropy Collaboration} et~al., 2018, \mn@doi [\aj] {10.3847/1538-3881/aabc4f},
  \href {https://ui.adsabs.harvard.edu/abs/2018AJ....156..123A} {156, 123}

\bibitem[\protect\citeauthoryear{{Balestra} et~al.,}{{Balestra}
  et~al.}{2010}]{bal10}
{Balestra} I.,  et~al., 2010, \mn@doi [\aap] {10.1051/0004-6361/200913626},
  \href {https://ui.adsabs.harvard.edu/abs/2010A&A...512A..12B} {512, A12}

\bibitem[\protect\citeauthoryear{{Balogh}, {Morris}, {Yee}, {Carlberg}  \&
  {Ellingson}}{{Balogh} et~al.}{1999}]{bal99}
{Balogh} M.~L.,  {Morris} S.~L.,  {Yee} H.~K.~C.,  {Carlberg} R.~G.,
  {Ellingson} E.,  1999, \mn@doi [\apj] {10.1086/308056}, \href
  {https://ui.adsabs.harvard.edu/abs/1999ApJ...527...54B} {527, 54}

\bibitem[\protect\citeauthoryear{{Barnes} \& {Hernquist}}{{Barnes} \&
  {Hernquist}}{1991}]{bar91}
{Barnes} J.~E.,  {Hernquist} L.~E.,  1991, \mn@doi [\apjl] {10.1086/185978},
  \href {https://ui.adsabs.harvard.edu/abs/1991ApJ...370L..65B} {370, L65}

\bibitem[\protect\citeauthoryear{{Barnes} \& {Hernquist}}{{Barnes} \&
  {Hernquist}}{1996}]{bar96}
{Barnes} J.~E.,  {Hernquist} L.,  1996, \mn@doi [\apj] {10.1086/177957}, \href
  {https://ui.adsabs.harvard.edu/abs/1996ApJ...471..115B} {471, 115}

\bibitem[\protect\citeauthoryear{{Belli}, {Newman}  \& {Ellis}}{{Belli}
  et~al.}{2017}]{bel17}
{Belli} S.,  {Newman} A.~B.,   {Ellis} R.~S.,  2017, \mn@doi [\apj]
  {10.3847/1538-4357/834/1/18}, \href
  {https://ui.adsabs.harvard.edu/abs/2017ApJ...834...18B} {834, 18}

\bibitem[\protect\citeauthoryear{{Belli}, {Newman}  \& {Ellis}}{{Belli}
  et~al.}{2019}]{bel19}
{Belli} S.,  {Newman} A.~B.,   {Ellis} R.~S.,  2019, \mn@doi [\apj]
  {10.3847/1538-4357/ab07af}, \href
  {https://ui.adsabs.harvard.edu/abs/2019ApJ...874...17B} {874, 17}

\bibitem[\protect\citeauthoryear{{Biggs} et~al.,}{{Biggs} et~al.}{2011}]{big11}
{Biggs} A.~D.,  et~al., 2011, \mn@doi [\mnras]
  {10.1111/j.1365-2966.2010.18132.x}, \href
  {https://ui.adsabs.harvard.edu/abs/2011MNRAS.413.2314B} {413, 2314}

\bibitem[\protect\citeauthoryear{{Blumenthal} \& {Barnes}}{{Blumenthal} \&
  {Barnes}}{2018}]{blu18}
{Blumenthal} K.~A.,  {Barnes} J.~E.,  2018, \mn@doi [\mnras]
  {10.1093/mnras/sty1605}, \href
  {https://ui.adsabs.harvard.edu/abs/2018MNRAS.479.3952B} {479, 3952}

\bibitem[\protect\citeauthoryear{{Brammer}, {van Dokkum}  \& {Coppi}}{{Brammer}
  et~al.}{2008}]{bra08}
{Brammer} G.~B.,  {van Dokkum} P.~G.,   {Coppi} P.,  2008, \mn@doi [\apj]
  {10.1086/591786}, \href
  {https://ui.adsabs.harvard.edu/abs/2008ApJ...686.1503B} {686, 1503}

\bibitem[\protect\citeauthoryear{{Calzetti}, {Armus}, {Bohlin}, {Kinney},
  {Koornneef}  \& {Storchi-Bergmann}}{{Calzetti} et~al.}{2000}]{cal00}
{Calzetti} D.,  {Armus} L.,  {Bohlin} R.~C.,  {Kinney} A.~L.,  {Koornneef} J.,
   {Storchi-Bergmann} T.,  2000, \mn@doi [\apj] {10.1086/308692}, \href
  {https://ui.adsabs.harvard.edu/abs/2000ApJ...533..682C} {533, 682}

\bibitem[\protect\citeauthoryear{{Cappellari}}{{Cappellari}}{2002}]{cap02}
{Cappellari} M.,  2002, \mn@doi [\mnras] {10.1046/j.1365-8711.2002.05412.x},
  \href {https://ui.adsabs.harvard.edu/abs/2002MNRAS.333..400C} {333, 400}

\bibitem[\protect\citeauthoryear{{Cappellari}}{{Cappellari}}{2008}]{cap08}
{Cappellari} M.,  2008, \mn@doi [\mnras] {10.1111/j.1365-2966.2008.13754.x},
  \href {https://ui.adsabs.harvard.edu/abs/2008MNRAS.390...71C} {390, 71}

\bibitem[\protect\citeauthoryear{{Cappellari}}{{Cappellari}}{2016}]{cap16}
{Cappellari} M.,  2016, \mn@doi [\araa] {10.1146/annurev-astro-082214-122432},
  \href {https://ui.adsabs.harvard.edu/abs/2016ARA&A..54..597C} {54, 597}

\bibitem[\protect\citeauthoryear{{Cappellari}}{{Cappellari}}{2017}]{cap17}
{Cappellari} M.,  2017, \mn@doi [\mnras] {10.1093/mnras/stw3020}, \href
  {https://ui.adsabs.harvard.edu/abs/2017MNRAS.466..798C} {466, 798}

\bibitem[\protect\citeauthoryear{{Cappellari} \& {Emsellem}}{{Cappellari} \&
  {Emsellem}}{2004}]{cap04}
{Cappellari} M.,  {Emsellem} E.,  2004, \mn@doi [\pasp] {10.1086/381875}, \href
  {https://ui.adsabs.harvard.edu/abs/2004PASP..116..138C} {116, 138}

\bibitem[\protect\citeauthoryear{{Cappellari} et~al.,}{{Cappellari}
  et~al.}{2006}]{cap06}
{Cappellari} M.,  et~al., 2006, \mn@doi [\mnras]
  {10.1111/j.1365-2966.2005.09981.x}, \href
  {https://ui.adsabs.harvard.edu/abs/2006MNRAS.366.1126C} {366, 1126}

\bibitem[\protect\citeauthoryear{{Cappellari} et~al.,}{{Cappellari}
  et~al.}{2013}]{cap13}
{Cappellari} M.,  et~al., 2013, \mn@doi [\mnras] {10.1093/mnras/stt562}, \href
  {https://ui.adsabs.harvard.edu/abs/2013MNRAS.432.1709C} {432, 1709}

\bibitem[\protect\citeauthoryear{{Chabrier}}{{Chabrier}}{2003}]{cha03}
{Chabrier} G.,  2003, \mn@doi [\pasp] {10.1086/376392}, \href
  {https://ui.adsabs.harvard.edu/abs/2003PASP..115..763C} {115, 763}

\bibitem[\protect\citeauthoryear{{Cibinel} et~al.,}{{Cibinel}
  et~al.}{2019}]{cib19}
{Cibinel} A.,  et~al., 2019, \mn@doi [\mnras] {10.1093/mnras/stz690}, \href
  {https://ui.adsabs.harvard.edu/abs/2019MNRAS.485.5631C} {485, 5631}

\bibitem[\protect\citeauthoryear{{Conroy} \& {Gunn}}{{Conroy} \&
  {Gunn}}{2010}]{con10}
{Conroy} C.,  {Gunn} J.~E.,  2010, \mn@doi [\apj]
  {10.1088/0004-637X/712/2/833}, \href
  {https://ui.adsabs.harvard.edu/abs/2010ApJ...712..833C} {712, 833}

\bibitem[\protect\citeauthoryear{{Cox}, {Jonsson}, {Somerville}, {Primack}  \&
  {Dekel}}{{Cox} et~al.}{2008}]{cox08}
{Cox} T.~J.,  {Jonsson} P.,  {Somerville} R.~S.,  {Primack} J.~R.,   {Dekel}
  A.,  2008, \mn@doi [\mnras] {10.1111/j.1365-2966.2007.12730.x}, \href
  {https://ui.adsabs.harvard.edu/abs/2008MNRAS.384..386C} {384, 386}

\bibitem[\protect\citeauthoryear{{Dai} et~al.,}{{Dai} et~al.}{2021}]{dai21}
{Dai} Y.~S.,  et~al., 2021, \mn@doi [\apj] {10.3847/1538-4357/ac2f96}, \href
  {https://ui.adsabs.harvard.edu/abs/2021ApJ...923..156D} {923, 156}

\bibitem[\protect\citeauthoryear{{Delhaize} et~al.,}{{Delhaize}
  et~al.}{2017}]{del17}
{Delhaize} J.,  et~al., 2017, \mn@doi [\aap] {10.1051/0004-6361/201629430},
  \href {https://ui.adsabs.harvard.edu/abs/2017A&A...602A...4D} {602, A4}

\bibitem[\protect\citeauthoryear{{Di Matteo}, {Combes}, {Melchior}  \&
  {Semelin}}{{Di Matteo} et~al.}{2007}]{di07}
{Di Matteo} P.,  {Combes} F.,  {Melchior} A.~L.,   {Semelin} B.,  2007, \mn@doi
  [\aap] {10.1051/0004-6361:20066959}, \href
  {https://ui.adsabs.harvard.edu/abs/2007A&A...468...61D} {468, 61}

\bibitem[\protect\citeauthoryear{{Di Matteo}, {Bournaud}, {Martig}, {Combes},
  {Melchior}  \& {Semelin}}{{Di Matteo} et~al.}{2008}]{di08}
{Di Matteo} P.,  {Bournaud} F.,  {Martig} M.,  {Combes} F.,  {Melchior} A.~L.,
   {Semelin} B.,  2008, \mn@doi [\aap] {10.1051/0004-6361:200809480}, \href
  {https://ui.adsabs.harvard.edu/abs/2008A&A...492...31D} {492, 31}

\bibitem[\protect\citeauthoryear{{Draine} \& {Li}}{{Draine} \&
  {Li}}{2007}]{dra07}
{Draine} B.~T.,  {Li} A.,  2007, \mn@doi [\apj] {10.1086/511055}, \href
  {https://ui.adsabs.harvard.edu/abs/2007ApJ...657..810D} {657, 810}

\bibitem[\protect\citeauthoryear{{Draine} et~al.,}{{Draine}
  et~al.}{2014}]{dra14}
{Draine} B.~T.,  et~al., 2014, \mn@doi [\apj] {10.1088/0004-637X/780/2/172},
  \href {https://ui.adsabs.harvard.edu/abs/2014ApJ...780..172D} {780, 172}

\bibitem[\protect\citeauthoryear{{Duncan} et~al.,}{{Duncan}
  et~al.}{2019}]{dun19}
{Duncan} K.,  et~al., 2019, \mn@doi [\apj] {10.3847/1538-4357/ab148a}, \href
  {https://ui.adsabs.harvard.edu/abs/2019ApJ...876..110D} {876, 110}

\bibitem[\protect\citeauthoryear{{Elbaz} et~al.,}{{Elbaz} et~al.}{2011}]{elb11}
{Elbaz} D.,  et~al., 2011, \mn@doi [\aap] {10.1051/0004-6361/201117239}, \href
  {https://ui.adsabs.harvard.edu/abs/2011A&A...533A.119E} {533, A119}

\bibitem[\protect\citeauthoryear{{Ellison}, {Patton}, {Simard}  \&
  {McConnachie}}{{Ellison} et~al.}{2008}]{ell08}
{Ellison} S.~L.,  {Patton} D.~R.,  {Simard} L.,   {McConnachie} A.~W.,  2008,
  \mn@doi [\aj] {10.1088/0004-6256/135/5/1877}, \href
  {https://ui.adsabs.harvard.edu/abs/2008AJ....135.1877E} {135, 1877}

\bibitem[\protect\citeauthoryear{{Esdaile} et~al.,}{{Esdaile}
  et~al.}{2021}]{esd21}
{Esdaile} J.,  et~al., 2021, \mn@doi [\apjl] {10.3847/2041-8213/abe11e}, \href
  {https://ui.adsabs.harvard.edu/abs/2021ApJ...908L..35E} {908, L35}

\bibitem[\protect\citeauthoryear{{Fensch} et~al.,}{{Fensch}
  et~al.}{2017}]{fen17}
{Fensch} J.,  et~al., 2017, \mn@doi [\mnras] {10.1093/mnras/stw2920}, \href
  {https://ui.adsabs.harvard.edu/abs/2017MNRAS.465.1934F} {465, 1934}

\bibitem[\protect\citeauthoryear{{Giavalisco} et~al.,}{{Giavalisco}
  et~al.}{2004}]{gia04}
{Giavalisco} M.,  et~al., 2004, \mn@doi [\apjl] {10.1086/379232}, \href
  {https://ui.adsabs.harvard.edu/abs/2004ApJ...600L..93G} {600, L93}

\bibitem[\protect\citeauthoryear{{Griffin} et~al.,}{{Griffin}
  et~al.}{2010}]{gri10}
{Griffin} M.~J.,  et~al., 2010, \mn@doi [\aap] {10.1051/0004-6361/201014519},
  \href {https://ui.adsabs.harvard.edu/abs/2010A&A...518L...3G} {518, L3}

\bibitem[\protect\citeauthoryear{{Grogin} et~al.,}{{Grogin}
  et~al.}{2011}]{gro11}
{Grogin} N.~A.,  et~al., 2011, \mn@doi [\apjs] {10.1088/0067-0049/197/2/35},
  \href {https://ui.adsabs.harvard.edu/abs/2011ApJS..197...35G} {197, 35}

\bibitem[\protect\citeauthoryear{Harris et~al.,}{Harris et~al.}{2020}]{har20}
Harris C.~R.,  et~al., 2020, \mn@doi [Nature] {10.1038/s41586-020-2649-2}, 585,
  357–362

\bibitem[\protect\citeauthoryear{{Hopkins}, {Hernquist}, {Cox}  \&
  {Kere{\v{s}}}}{{Hopkins} et~al.}{2008}]{hop08}
{Hopkins} P.~F.,  {Hernquist} L.,  {Cox} T.~J.,   {Kere{\v{s}}} D.,  2008,
  \mn@doi [\apjs] {10.1086/524362}, \href
  {https://ui.adsabs.harvard.edu/abs/2008ApJS..175..356H} {175, 356}

\bibitem[\protect\citeauthoryear{{Hopkins} et~al.,}{{Hopkins}
  et~al.}{2010}]{hop10}
{Hopkins} P.~F.,  et~al., 2010, \mn@doi [\apj] {10.1088/0004-637X/724/2/915},
  \href {https://ui.adsabs.harvard.edu/abs/2010ApJ...724..915H} {724, 915}

\bibitem[\protect\citeauthoryear{{Hopkins}, {Quataert}  \& {Murray}}{{Hopkins}
  et~al.}{2011}]{hop11}
{Hopkins} P.~F.,  {Quataert} E.,   {Murray} N.,  2011, \mn@doi [\mnras]
  {10.1111/j.1365-2966.2011.19306.x}, \href
  {https://ui.adsabs.harvard.edu/abs/2011MNRAS.417..950H} {417, 950}

\bibitem[\protect\citeauthoryear{{Hopkins}, {Kere{\v{s}}}, {O{\~n}orbe},
  {Faucher-Gigu{\`e}re}, {Quataert}, {Murray}  \& {Bullock}}{{Hopkins}
  et~al.}{2014}]{hop14}
{Hopkins} P.~F.,  {Kere{\v{s}}} D.,  {O{\~n}orbe} J.,  {Faucher-Gigu{\`e}re}
  C.-A.,  {Quataert} E.,  {Murray} N.,   {Bullock} J.~S.,  2014, \mn@doi
  [\mnras] {10.1093/mnras/stu1738}, \href
  {https://ui.adsabs.harvard.edu/abs/2014MNRAS.445..581H} {445, 581}

\bibitem[\protect\citeauthoryear{{Hopkins} et~al.,}{{Hopkins}
  et~al.}{2018}]{hop18}
{Hopkins} P.~F.,  et~al., 2018, \mn@doi [\mnras] {10.1093/mnras/sty1690}, \href
  {https://ui.adsabs.harvard.edu/abs/2018MNRAS.480..800H} {480, 800}

\bibitem[\protect\citeauthoryear{{Horstman} et~al.,}{{Horstman}
  et~al.}{2021}]{hor21}
{Horstman} K.,  et~al., 2021, \mn@doi [\mnras] {10.1093/mnras/staa3502}, \href
  {https://ui.adsabs.harvard.edu/abs/2021MNRAS.501..137H} {501, 137}

\bibitem[\protect\citeauthoryear{{Hunter}}{{Hunter}}{2007}]{hun07}
{Hunter} J.~D.,  2007, Computing in Science \& Engineering, 9, 90

\bibitem[\protect\citeauthoryear{{Jin} et~al.,}{{Jin} et~al.}{2018}]{jin18}
{Jin} S.,  et~al., 2018, \mn@doi [\apj] {10.3847/1538-4357/aad4af}, \href
  {https://ui.adsabs.harvard.edu/abs/2018ApJ...864...56J} {864, 56}

\bibitem[\protect\citeauthoryear{Johnson et~al.,}{Johnson
  et~al.}{2021a}]{joh21}
Johnson B.,  et~al., 2021a, dfm/python-fsps: python-fsps v0.4.1rc1,
  \mn@doi{10.5281/zenodo.4737461}, \url
  {https://doi.org/10.5281/zenodo.4737461}

\bibitem[\protect\citeauthoryear{{Johnson}, {Leja}, {Conroy}  \&
  {Speagle}}{{Johnson} et~al.}{2021b}]{joh21prospector}
{Johnson} B.~D.,  {Leja} J.,  {Conroy} C.,   {Speagle} J.~S.,  2021b, \mn@doi
  [\apjs] {10.3847/1538-4365/abef67}, \href
  {https://ui.adsabs.harvard.edu/abs/2021ApJS..254...22J} {254, 22}

\bibitem[\protect\citeauthoryear{{Kartaltepe} et~al.,}{{Kartaltepe}
  et~al.}{2015}]{kar15}
{Kartaltepe} J.~S.,  et~al., 2015, \mn@doi [\apjs]
  {10.1088/0067-0049/221/1/11}, \href
  {https://ui.adsabs.harvard.edu/abs/2015ApJS..221...11K} {221, 11}

\bibitem[\protect\citeauthoryear{{Koekemoer} et~al.,}{{Koekemoer}
  et~al.}{2011}]{koe11}
{Koekemoer} A.~M.,  et~al., 2011, \mn@doi [\apjs] {10.1088/0067-0049/197/2/36},
  \href {https://ui.adsabs.harvard.edu/abs/2011ApJS..197...36K} {197, 36}

\bibitem[\protect\citeauthoryear{{Kokorev} et~al.,}{{Kokorev}
  et~al.}{2021}]{kok21}
{Kokorev} V.~I.,  et~al., 2021, \mn@doi [\apj] {10.3847/1538-4357/ac18ce},
  \href {https://ui.adsabs.harvard.edu/abs/2021ApJ...921...40K} {921, 40}

\bibitem[\protect\citeauthoryear{{Kormendy} \& {Ho}}{{Kormendy} \&
  {Ho}}{2013}]{kor13}
{Kormendy} J.,  {Ho} L.~C.,  2013, \mn@doi [\araa]
  {10.1146/annurev-astro-082708-101811}, \href
  {https://ui.adsabs.harvard.edu/abs/2013ARA&A..51..511K} {51, 511}

\bibitem[\protect\citeauthoryear{{Kriek} et~al.,}{{Kriek} et~al.}{2008}]{kri08}
{Kriek} M.,  et~al., 2008, \mn@doi [\apj] {10.1086/528945}, \href
  {https://ui.adsabs.harvard.edu/abs/2008ApJ...677..219K} {677, 219}

\bibitem[\protect\citeauthoryear{{Kriek}, {van Dokkum}, {Labb{\'e}}, {Franx},
  {Illingworth}, {Marchesini}  \& {Quadri}}{{Kriek} et~al.}{2009}]{kri09}
{Kriek} M.,  {van Dokkum} P.~G.,  {Labb{\'e}} I.,  {Franx} M.,  {Illingworth}
  G.~D.,  {Marchesini} D.,   {Quadri} R.~F.,  2009, \mn@doi [\apj]
  {10.1088/0004-637X/700/1/221}, \href
  {https://ui.adsabs.harvard.edu/abs/2009ApJ...700..221K} {700, 221}

\bibitem[\protect\citeauthoryear{{Kriek} et~al.,}{{Kriek} et~al.}{2015}]{kri15}
{Kriek} M.,  et~al., 2015, \mn@doi [\apjs] {10.1088/0067-0049/218/2/15}, \href
  {https://ui.adsabs.harvard.edu/abs/2015ApJS..218...15K} {218, 15}

\bibitem[\protect\citeauthoryear{{Kriek} et~al.,}{{Kriek} et~al.}{2016}]{kri16}
{Kriek} M.,  et~al., 2016, \mn@doi [\nat] {10.1038/nature20570}, \href
  {https://ui.adsabs.harvard.edu/abs/2016Natur.540..248K} {540, 248}

\bibitem[\protect\citeauthoryear{{Lofthouse}, {Kaviraj}, {Conselice},
  {Mortlock}  \& {Hartley}}{{Lofthouse} et~al.}{2017}]{lof17}
{Lofthouse} E.~K.,  {Kaviraj} S.,  {Conselice} C.~J.,  {Mortlock} A.,
  {Hartley} W.,  2017, \mn@doi [\mnras] {10.1093/mnras/stw2895}, \href
  {https://ui.adsabs.harvard.edu/abs/2017MNRAS.465.2895L} {465, 2895}

\bibitem[\protect\citeauthoryear{{Lotz}, {Jonsson}, {Cox}, {Croton}, {Primack},
  {Somerville}  \& {Stewart}}{{Lotz} et~al.}{2011}]{lot11}
{Lotz} J.~M.,  {Jonsson} P.,  {Cox} T.~J.,  {Croton} D.,  {Primack} J.~R.,
  {Somerville} R.~S.,   {Stewart} K.,  2011, \mn@doi [\apj]
  {10.1088/0004-637X/742/2/103}, \href
  {https://ui.adsabs.harvard.edu/abs/2011ApJ...742..103L} {742, 103}

\bibitem[\protect\citeauthoryear{{Luo} et~al.,}{{Luo} et~al.}{2008}]{luo08}
{Luo} B.,  et~al., 2008, \mn@doi [\apjs] {10.1086/591248}, \href
  {https://ui.adsabs.harvard.edu/abs/2008ApJS..179...19L} {179, 19}

\bibitem[\protect\citeauthoryear{{Man}, {Toft}, {Zirm}, {Wuyts}  \& {van der
  Wel}}{{Man} et~al.}{2012}]{man12}
{Man} A. W.~S.,  {Toft} S.,  {Zirm} A.~W.,  {Wuyts} S.,   {van der Wel} A.,
  2012, \mn@doi [\apj] {10.1088/0004-637X/744/2/85}, \href
  {https://ui.adsabs.harvard.edu/abs/2012ApJ...744...85M} {744, 85}

\bibitem[\protect\citeauthoryear{{Man}, {Zirm}  \& {Toft}}{{Man}
  et~al.}{2016}]{man16}
{Man} A. W.~S.,  {Zirm} A.~W.,   {Toft} S.,  2016, \mn@doi [\apj]
  {10.3847/0004-637X/830/2/89}, \href
  {https://ui.adsabs.harvard.edu/abs/2016ApJ...830...89M} {830, 89}

\bibitem[\protect\citeauthoryear{{Mantha} et~al.,}{{Mantha}
  et~al.}{2018}]{man18}
{Mantha} K.~B.,  et~al., 2018, \mn@doi [\mnras] {10.1093/mnras/stx3260}, \href
  {https://ui.adsabs.harvard.edu/abs/2018MNRAS.475.1549M} {475, 1549}

\bibitem[\protect\citeauthoryear{{McLean} et~al.,}{{McLean}
  et~al.}{2012}]{mcl12}
{McLean} I.~S.,  et~al., 2012, in Ground-based and Airborne Instrumentation for
  Astronomy IV. p. 84460J, \mn@doi{10.1117/12.924794}

\bibitem[\protect\citeauthoryear{{Miller} et~al.,}{{Miller}
  et~al.}{2013}]{mil13}
{Miller} N.~A.,  et~al., 2013, \mn@doi [\apjs] {10.1088/0067-0049/205/2/13},
  \href {https://ui.adsabs.harvard.edu/abs/2013ApJS..205...13M} {205, 13}

\bibitem[\protect\citeauthoryear{{Millman} \& {Aivazis}}{{Millman} \&
  {Aivazis}}{2011}]{mil11}
{Millman} K.~J.,  {Aivazis} M.,  2011, Computing in Science \& Engineering, 13,
  9

\bibitem[\protect\citeauthoryear{{Momcheva} et~al.,}{{Momcheva}
  et~al.}{2016}]{mom16}
{Momcheva} I.~G.,  et~al., 2016, \mn@doi [\apjs] {10.3847/0067-0049/225/2/27},
  \href {https://ui.adsabs.harvard.edu/abs/2016ApJS..225...27M} {225, 27}

\bibitem[\protect\citeauthoryear{{Moreno}, {Torrey}, {Ellison}, {Patton},
  {Bluck}, {Bansal}  \& {Hernquist}}{{Moreno} et~al.}{2015}]{mor15}
{Moreno} J.,  {Torrey} P.,  {Ellison} S.~L.,  {Patton} D.~R.,  {Bluck} A.
  F.~L.,  {Bansal} G.,   {Hernquist} L.,  2015, \mn@doi [\mnras]
  {10.1093/mnras/stv094}, \href
  {https://ui.adsabs.harvard.edu/abs/2015MNRAS.448.1107M} {448, 1107}

\bibitem[\protect\citeauthoryear{{Moreno} et~al.,}{{Moreno}
  et~al.}{2019}]{mor19}
{Moreno} J.,  et~al., 2019, \mn@doi [\mnras] {10.1093/mnras/stz417}, \href
  {https://ui.adsabs.harvard.edu/abs/2019MNRAS.485.1320M} {485, 1320}

\bibitem[\protect\citeauthoryear{{Mullaney}, {Alexander}, {Goulding}  \&
  {Hickox}}{{Mullaney} et~al.}{2011}]{mul11}
{Mullaney} J.~R.,  {Alexander} D.~M.,  {Goulding} A.~D.,   {Hickox} R.~C.,
  2011, \mn@doi [\mnras] {10.1111/j.1365-2966.2011.18448.x}, \href
  {https://ui.adsabs.harvard.edu/abs/2011MNRAS.414.1082M} {414, 1082}

\bibitem[\protect\citeauthoryear{{Muzzin}, {Marchesini}, {van Dokkum},
  {Labb{\'e}}, {Kriek}  \& {Franx}}{{Muzzin} et~al.}{2009}]{muz09}
{Muzzin} A.,  {Marchesini} D.,  {van Dokkum} P.~G.,  {Labb{\'e}} I.,  {Kriek}
  M.,   {Franx} M.,  2009, \mn@doi [\apj] {10.1088/0004-637X/701/2/1839}, \href
  {https://ui.adsabs.harvard.edu/abs/2009ApJ...701.1839M} {701, 1839}

\bibitem[\protect\citeauthoryear{{Newman}, {Belli}, {Ellis}  \&
  {Patel}}{{Newman} et~al.}{2018a}]{new18a}
{Newman} A.~B.,  {Belli} S.,  {Ellis} R.~S.,   {Patel} S.~G.,  2018a, \mn@doi
  [\apj] {10.3847/1538-4357/aacd4d}, \href
  {https://ui.adsabs.harvard.edu/abs/2018ApJ...862..125N} {862, 125}

\bibitem[\protect\citeauthoryear{{Newman}, {Belli}, {Ellis}  \&
  {Patel}}{{Newman} et~al.}{2018b}]{new18b}
{Newman} A.~B.,  {Belli} S.,  {Ellis} R.~S.,   {Patel} S.~G.,  2018b, \mn@doi
  [\apj] {10.3847/1538-4357/aacd4f}, \href
  {https://ui.adsabs.harvard.edu/abs/2018ApJ...862..126N} {862, 126}

\bibitem[\protect\citeauthoryear{{Oliphant}}{{Oliphant}}{2007}]{oli07}
{Oliphant} T.~E.,  2007, Computing in Science \& Engineering, 9, 10

\bibitem[\protect\citeauthoryear{{Patton}, {Ellison}, {Simard}, {McConnachie}
  \& {Mendel}}{{Patton} et~al.}{2011}]{pat11}
{Patton} D.~R.,  {Ellison} S.~L.,  {Simard} L.,  {McConnachie} A.~W.,
  {Mendel} J.~T.,  2011, \mn@doi [\mnras] {10.1111/j.1365-2966.2010.17932.x},
  \href {https://ui.adsabs.harvard.edu/abs/2011MNRAS.412..591P} {412, 591}

\bibitem[\protect\citeauthoryear{{Patton}, {Torrey}, {Ellison}, {Mendel}  \&
  {Scudder}}{{Patton} et~al.}{2013}]{pat13}
{Patton} D.~R.,  {Torrey} P.,  {Ellison} S.~L.,  {Mendel} J.~T.,   {Scudder}
  J.~M.,  2013, \mn@doi [\mnras] {10.1093/mnrasl/slt058}, \href
  {https://ui.adsabs.harvard.edu/abs/2013MNRAS.433L..59P} {433, L59}

\bibitem[\protect\citeauthoryear{{Perez} \& {Granger}}{{Perez} \&
  {Granger}}{2007}]{per07}
{Perez} F.,  {Granger} B.~E.,  2007, Computing in Science \& Engineering, 9, 21

\bibitem[\protect\citeauthoryear{{Perley}, {Chandler}, {Butler}  \&
  {Wrobel}}{{Perley} et~al.}{2011}]{per11}
{Perley} R.~A.,  {Chandler} C.~J.,  {Butler} B.~J.,   {Wrobel} J.~M.,  2011,
  \mn@doi [\apjl] {10.1088/2041-8205/739/1/L1}, \href
  {https://ui.adsabs.harvard.edu/abs/2011ApJ...739L...1P} {739, L1}

\bibitem[\protect\citeauthoryear{{Poglitsch} et~al.,}{{Poglitsch}
  et~al.}{2010}]{pog10}
{Poglitsch} A.,  et~al., 2010, \mn@doi [\aap] {10.1051/0004-6361/201014535},
  \href {https://ui.adsabs.harvard.edu/abs/2010A&A...518L...2P} {518, L2}

\bibitem[\protect\citeauthoryear{{Rieke} et~al.,}{{Rieke} et~al.}{2004}]{rie04}
{Rieke} G.~H.,  et~al., 2004, \mn@doi [\apjs] {10.1086/422717}, \href
  {https://ui.adsabs.harvard.edu/abs/2004ApJS..154...25R} {154, 25}

\bibitem[\protect\citeauthoryear{{Schawinski}, {Treister}, {Urry}, {Cardamone},
  {Simmons}  \& {Yi}}{{Schawinski} et~al.}{2011}]{sch11}
{Schawinski} K.,  {Treister} E.,  {Urry} C.~M.,  {Cardamone} C.~N.,  {Simmons}
  B.,   {Yi} S.~K.,  2011, \mn@doi [\apjl] {10.1088/2041-8205/727/2/L31}, \href
  {https://ui.adsabs.harvard.edu/abs/2011ApJ...727L..31S} {727, L31}

\bibitem[\protect\citeauthoryear{{Scudder}, {Ellison}, {Torrey}, {Patton}  \&
  {Mendel}}{{Scudder} et~al.}{2012}]{scu12}
{Scudder} J.~M.,  {Ellison} S.~L.,  {Torrey} P.,  {Patton} D.~R.,   {Mendel}
  J.~T.,  2012, \mn@doi [\mnras] {10.1111/j.1365-2966.2012.21749.x}, \href
  {https://ui.adsabs.harvard.edu/abs/2012MNRAS.426..549S} {426, 549}

\bibitem[\protect\citeauthoryear{{Scudder}, {Ellison}, {Momjian}, {Rosenberg},
  {Torrey}, {Patton}, {Fertig}  \& {Mendel}}{{Scudder} et~al.}{2015}]{scu15}
{Scudder} J.~M.,  {Ellison} S.~L.,  {Momjian} E.,  {Rosenberg} J.~L.,  {Torrey}
  P.,  {Patton} D.~R.,  {Fertig} D.,   {Mendel} J.~T.,  2015, \mn@doi [\mnras]
  {10.1093/mnras/stv588}, \href
  {https://ui.adsabs.harvard.edu/abs/2015MNRAS.449.3719S} {449, 3719}

\bibitem[\protect\citeauthoryear{{Skelton} et~al.,}{{Skelton}
  et~al.}{2014}]{ske14}
{Skelton} R.~E.,  et~al., 2014, \mn@doi [\apjs] {10.1088/0067-0049/214/2/24},
  \href {https://ui.adsabs.harvard.edu/abs/2014ApJS..214...24S} {214, 24}

\bibitem[\protect\citeauthoryear{{Smail}, {Morrison}, {Gray}, {Owen}, {Ivison},
  {Kneib}  \& {Ellis}}{{Smail} et~al.}{1999}]{sma99}
{Smail} I.,  {Morrison} G.,  {Gray} M.~E.,  {Owen} F.~N.,  {Ivison} R.~J.,
  {Kneib} J.~P.,   {Ellis} R.~S.,  1999, \mn@doi [\apj] {10.1086/307934}, \href
  {https://ui.adsabs.harvard.edu/abs/1999ApJ...525..609S} {525, 609}

\bibitem[\protect\citeauthoryear{{Tacchella} et~al.,}{{Tacchella}
  et~al.}{2022}]{tac22}
{Tacchella} S.,  et~al., 2022, \mn@doi [\apj] {10.3847/1538-4357/ac4cad}, \href
  {https://ui.adsabs.harvard.edu/abs/2022ApJ...927..170T} {927, 170}

\bibitem[\protect\citeauthoryear{{Tanaka} et~al.,}{{Tanaka}
  et~al.}{2019}]{tan19}
{Tanaka} M.,  et~al., 2019, \mn@doi [\apjl] {10.3847/2041-8213/ab4ff3}, \href
  {https://ui.adsabs.harvard.edu/abs/2019ApJ...885L..34T} {885, L34}

\bibitem[\protect\citeauthoryear{{Tasca} et~al.,}{{Tasca} et~al.}{2014}]{tas14}
{Tasca} L.~A.~M.,  et~al., 2014, \mn@doi [\aap] {10.1051/0004-6361/201321507},
  \href {https://ui.adsabs.harvard.edu/abs/2014A&A...565A..10T} {565, A10}

\bibitem[\protect\citeauthoryear{{The pandas development team}}{{The pandas
  development team}}{2020}]{pandas20}
{The pandas development team} 2020, pandas-dev/pandas: Pandas,
  \mn@doi{10.5281/zenodo.3509134}, \url
  {https://doi.org/10.5281/zenodo.3509134}

\bibitem[\protect\citeauthoryear{{Tody}}{{Tody}}{1986}]{tod86}
{Tody} D.,  1986, in {Crawford} D.~L.,  ed.,  Society of Photo-Optical
  Instrumentation Engineers (SPIE) Conference Series Vol. 627, Instrumentation
  in astronomy VI. p.~733, \mn@doi{10.1117/12.968154}

\bibitem[\protect\citeauthoryear{{Tody}}{{Tody}}{1993}]{tod93}
{Tody} D.,  1993, in {Hanisch} R.~J.,  {Brissenden} R.~J.~V.,   {Barnes} J.,
  eds,  Astronomical Society of the Pacific Conference Series Vol. 52,
  Astronomical Data Analysis Software and Systems II. p.~173

\bibitem[\protect\citeauthoryear{{Toft}, {Gallazzi}, {Zirm}, {Wold}, {Zibetti},
  {Grillo}  \& {Man}}{{Toft} et~al.}{2012}]{tof12}
{Toft} S.,  {Gallazzi} A.,  {Zirm} A.,  {Wold} M.,  {Zibetti} S.,  {Grillo} C.,
    {Man} A.,  2012, \mn@doi [\apj] {10.1088/0004-637X/754/1/3}, \href
  {https://ui.adsabs.harvard.edu/abs/2012ApJ...754....3T} {754, 3}

\bibitem[\protect\citeauthoryear{{Toft} et~al.,}{{Toft} et~al.}{2017}]{tof17}
{Toft} S.,  et~al., 2017, \mn@doi [\nat] {10.1038/nature22388}, \href
  {https://ui.adsabs.harvard.edu/abs/2017Natur.546..510T} {546, 510}

\bibitem[\protect\citeauthoryear{{Vazdekis}, {S{\'a}nchez-Bl{\'a}zquez},
  {Falc{\'o}n-Barroso}, {Cenarro}, {Beasley}, {Cardiel}, {Gorgas}  \&
  {Peletier}}{{Vazdekis} et~al.}{2010}]{vaz10}
{Vazdekis} A.,  {S{\'a}nchez-Bl{\'a}zquez} P.,  {Falc{\'o}n-Barroso} J.,
  {Cenarro} A.~J.,  {Beasley} M.~A.,  {Cardiel} N.,  {Gorgas} J.,   {Peletier}
  R.~F.,  2010, \mn@doi [\mnras] {10.1111/j.1365-2966.2010.16407.x}, \href
  {https://ui.adsabs.harvard.edu/abs/2010MNRAS.404.1639V} {404, 1639}

\bibitem[\protect\citeauthoryear{{Ventou} et~al.,}{{Ventou}
  et~al.}{2017}]{ven17}
{Ventou} E.,  et~al., 2017, \mn@doi [\aap] {10.1051/0004-6361/201731586}, \href
  {https://ui.adsabs.harvard.edu/abs/2017A&A...608A...9V} {608, A9}

\bibitem[\protect\citeauthoryear{{Virtanen} et~al.,}{{Virtanen}
  et~al.}{2020}]{vir20}
{Virtanen} P.,  et~al., 2020, \mn@doi [Nature Methods]
  {10.1038/s41592-019-0686-2}, \href
  {https://ui.adsabs.harvard.edu/abs/2020NatMe..17..261V} {17, 261}

\bibitem[\protect\citeauthoryear{{Werner} et~al.,}{{Werner}
  et~al.}{2004}]{wer04}
{Werner} M.~W.,  et~al., 2004, \mn@doi [\apjs] {10.1086/422992}, \href
  {https://ui.adsabs.harvard.edu/abs/2004ApJS..154....1W} {154, 1}

\bibitem[\protect\citeauthoryear{{W}es {M}c{K}inney}{{W}es
  {M}c{K}inney}{2010}]{mckinney-proc-scipy-2010}
{W}es {M}c{K}inney 2010, in {S}t\'efan van~der {W}alt {J}arrod {M}illman eds,
  {P}roceedings of the 9th {P}ython in {S}cience {C}onference. pp 56 -- 61,
  \mn@doi{10.25080/Majora-92bf1922-00a}

\bibitem[\protect\citeauthoryear{{Whitaker}, {Kriek}, {van Dokkum}, {Bezanson},
  {Brammer}, {Franx}  \& {Labb{\'e}}}{{Whitaker} et~al.}{2012}]{whi12}
{Whitaker} K.~E.,  {Kriek} M.,  {van Dokkum} P.~G.,  {Bezanson} R.,  {Brammer}
  G.,  {Franx} M.,   {Labb{\'e}} I.,  2012, \mn@doi [\apj]
  {10.1088/0004-637X/745/2/179}, \href
  {https://ui.adsabs.harvard.edu/abs/2012ApJ...745..179W} {745, 179}

\bibitem[\protect\citeauthoryear{{Williams}, {Quadri}, {Franx}, {van Dokkum}
  \& {Labb{\'e}}}{{Williams} et~al.}{2009}]{wil09}
{Williams} R.~J.,  {Quadri} R.~F.,  {Franx} M.,  {van Dokkum} P.,   {Labb{\'e}}
  I.,  2009, \mn@doi [\apj] {10.1088/0004-637X/691/2/1879}, \href
  {https://ui.adsabs.harvard.edu/abs/2009ApJ...691.1879W} {691, 1879}

\bibitem[\protect\citeauthoryear{{Williams}, {Quadri}  \& {Franx}}{{Williams}
  et~al.}{2011}]{wil11}
{Williams} R.~J.,  {Quadri} R.~F.,   {Franx} M.,  2011, \mn@doi [\apjl]
  {10.1088/2041-8205/738/2/L25}, \href
  {https://ui.adsabs.harvard.edu/abs/2011ApJ...738L..25W} {738, L25}

\bibitem[\protect\citeauthoryear{{Wilson} et~al.,}{{Wilson}
  et~al.}{2019}]{wil19}
{Wilson} T.~J.,  et~al., 2019, \mn@doi [\apj] {10.3847/1538-4357/ab06ee}, \href
  {https://ui.adsabs.harvard.edu/abs/2019ApJ...874...18W} {874, 18}

\bibitem[\protect\citeauthoryear{{Worthey} \& {Ottaviani}}{{Worthey} \&
  {Ottaviani}}{1997}]{wor97}
{Worthey} G.,  {Ottaviani} D.~L.,  1997, \mn@doi [\apjs] {10.1086/313021},
  \href {https://ui.adsabs.harvard.edu/abs/1997ApJS..111..377W} {111, 377}

\bibitem[\protect\citeauthoryear{{Yano}, {Kriek}, {van der Wel}  \&
  {Whitaker}}{{Yano} et~al.}{2016}]{yan16}
{Yano} M.,  {Kriek} M.,  {van der Wel} A.,   {Whitaker} K.~E.,  2016, \mn@doi
  [\apjl] {10.3847/2041-8205/817/2/L21}, \href
  {https://ui.adsabs.harvard.edu/abs/2016ApJ...817L..21Y} {817, L21}

\bibitem[\protect\citeauthoryear{{Zick} et~al.,}{{Zick} et~al.}{2018}]{zic18}
{Zick} T.~O.,  et~al., 2018, \mn@doi [\apjl] {10.3847/2041-8213/aae887}, \href
  {https://ui.adsabs.harvard.edu/abs/2018ApJ...867L..16Z} {867, L16}

\bibitem[\protect\citeauthoryear{{van Dokkum} \& {Brammer}}{{van Dokkum} \&
  {Brammer}}{2010}]{van10}
{van Dokkum} P.~G.,  {Brammer} G.,  2010, \mn@doi [\apjl]
  {10.1088/2041-8205/718/2/L73}, \href
  {https://ui.adsabs.harvard.edu/abs/2010ApJ...718L..73V} {718, L73}

\bibitem[\protect\citeauthoryear{{van de Sande} et~al.,}{{van de Sande}
  et~al.}{2013}]{van13}
{van de Sande} J.,  et~al., 2013, \mn@doi [\apj] {10.1088/0004-637X/771/2/85},
  \href {https://ui.adsabs.harvard.edu/abs/2013ApJ...771...85V} {771, 85}

\bibitem[\protect\citeauthoryear{{van der Walt}, {Colbert}  \&
  {Varoquaux}}{{van der Walt} et~al.}{2011}]{van11}
{van der Walt} S.,  {Colbert} S.~C.,   {Varoquaux} G.,  2011, Computing in
  Science \& Engineering, 13, 22

\bibitem[\protect\citeauthoryear{{van der Wel} et~al.,}{{van der Wel}
  et~al.}{2012}]{van12}
{van der Wel} A.,  et~al., 2012, \mn@doi [\apjs] {10.1088/0067-0049/203/2/24},
  \href {https://ui.adsabs.harvard.edu/abs/2012ApJS..203...24V} {203, 24}

\bibitem[\protect\citeauthoryear{{van der Wel} et~al.,}{{van der Wel}
  et~al.}{2014}]{van14}
{van der Wel} A.,  et~al., 2014, \mn@doi [\apj] {10.1088/0004-637X/788/1/28},
  \href {https://ui.adsabs.harvard.edu/abs/2014ApJ...788...28V} {788, 28}

\makeatother
\end{thebibliography}

\bsp
\label{lastpage}
\end{document}